\documentclass[usenatbib]{mnras}
\usepackage{mathptmx}
\usepackage[T1]{fontenc}
\usepackage{ae,aecompl}

\usepackage{graphicx}
\usepackage{amsmath}
\usepackage{wasysym}
\usepackage{xspace}
\usepackage{color}
\usepackage[inline]{enumitem}
\usepackage{multirow}
\usepackage{lineno}
\usepackage{threeparttable}
\usepackage{hyperref}

\newenvironment{tightcenter}{%
  \setlength\topsep{0pt}
  \setlength\parskip{0pt}
  \begin{center}
}{%
  \end{center}
}

\newcommand{\e}[1]{\ensuremath{\times 10^{#1}}}
\newcommand{\LCDM}{$\Lambda$CDM}

\newcommand{\TODO}[1]{}
\renewcommand{\TODO}[1]{{\color{red} TODO: {#1}}}

\newcommand{\re}[1]{{Eq.~\ref{eqn:#1}}}

\newcommand{\beq}{\begin{equation}} 
\newcommand{\eeq}{\end{equation}} 
\newcommand{\bea}{\begin{eqnarray}} 
\newcommand{\eea}{\end{eqnarray}}  
\newcommand{\bean}{\begin{eqnarray*}} 
\newcommand{\eean}{\end{eqnarray*}} 
\newcommand{\bal}{\begin{align}} 
\newcommand{\eal}{\end{align}} 
\newcommand{\bfk}{{\bf k}}
\newcommand{\bfx}{{\bf x}}
\newcommand{\bfu}{{\bf u}}
\newcommand{\bfv}{{\bf v}}
\newcommand{\bfr}{{\bf r}}
\newcommand{\bfR}{{\bf R}}
\newcommand{\bfe}{{\bf e}}
\newcommand{\al}{\alpha}
\newcommand{\bfF}{{\bf F}}
\newcommand{\bfg}{{\bf g}}
\newcommand{\bfS}{{\bf S}}
\newcommand{\kNy}{{k_{\rm Nyquist}}}

\title[Improving Cosmological $N$-body ICs]{Improving Initial Conditions for Cosmological $N$-Body Simulations}

\author[Garrison et al.]{Lehman H.~Garrison,$^1$\thanks{E-mail: \texttt{lgarrison@cfa.harvard.edu}}
Daniel J.~Eisenstein,$^1$
Douglas Ferrer,$^1$
Marc V.~Metchnik,$^2$
\newauthor
and Philip A.~Pinto$^2$
\\
$^1$Harvard-Smithsonian Center for Astrophysics, 60 Garden St., Cambridge, MA 02138\\
$^2$Steward Observatory, University of Arizona, 933 N. Cherry Ave., Tucson, AZ 85121
}

\date{Accepted XXX. Received YYY; in original form ZZZ}

\pubyear{2016}

\begin{document}
\label{firstpage}
\pagerange{\pageref{firstpage}--\pageref{lastpage}}
\maketitle

\begin{abstract}
In cosmological $N$-body simulations, the representation of dark matter as discrete ``macroparticles'' suppresses the growth of structure, such that simulations no longer reproduce linear theory on small scales near $k_{\rm Nyquist}$.  \citeauthor{Marcos+2006} demonstrate that this is due to sparse sampling of modes near $k_{\rm Nyquist}$ and that the often-assumed continuum growing modes are not proper growing modes of the particle system.  We develop initial conditions that respect the particle linear theory growing modes and then rescale the mode amplitudes to account for growth suppression.  These ICs also allow us to take advantage of our very accurate $N$-body code \textsc{Abacus} to implement 2LPT in configuration space.  The combination of 2LPT and rescaling improves the accuracy of the late-time power spectra, halo mass functions, and halo clustering.  In particular, we achieve 1\% accuracy in the power spectrum down to $\kNy$, versus $\kNy/4$ without rescaling or $\kNy/13$ without 2LPT, relative to an oversampled reference simulation.  We anticipate that our 2LPT will be useful for large simulations where FFTs are expensive and that rescaling will be useful for suites of medium-resolution simulations used in cosmic emulators and galaxy survey mock catalogs.  Code to generate initial conditions is available at \url{https://github.com/lgarrison/zeldovich-PLT}.
\end{abstract}

\begin{keywords}
large-scale structure of Universe -- galaxies: haloes -- methods: numerical
\end{keywords}

\section{Introduction}
Cosmological $N$-body simulations are the state-of-the-art tool for predicting dark matter halo clustering and masses for a given cosmology.  In most cosmological models, a large fraction of mass is in the form of dark matter and thus behaves as a collisionless ``fluid'' well described by the coupled Vlasov-Poisson equations.  $N$-body simulations take a discrete ``macroparticle'' sampling of this underlying fluid and then treat the evolution of the particle system as tracing the evolution of the fluid system \citep[for an alternative phase-space formulation of this problem, see][]{Hahn+2013}.  The applicability of results derived from $N$-body simulations thus assumes correspondence between the particle system and the fluid system.

This assumption has a number of well-documented violations (collectively known as ``discreteness effects'') at very early and very late times, such as correlations induced by the initial particle sampling \citep[see][]{Joyce_Marcos_2007a}, and two-body relaxation \citep[e.g.,][]{Binney_Knebe_2002,El-Zant_2006}.  A third, intermediate regime has received less attention, however: the evolution of the $N$-body system from its initial configuration up to the mildly non-linear regime.  The work of \cite{Marcos+2006} showed that in this regime, the assumptions of fluid linear theory are strongly violated on small scales.  Based on their work, we seek to correct these small-scale effects by modifying our initial conditions to respect the proper growing modes of the simulation and compensate for missing growth.  This directly addresses the improper growth of modes on small scales that \cite{Warren_2013} identified as the dominant systematic error in precision halo mass functions.

The underlying theory, developed by \cite{Marcos+2006}, is called \textit{particle linear theory} (PLT).  PLT is an analytical description of the evolution of a grid-like particle system that self-interacts under a $1/r^2$ force law.  It is a perturbative solution to the fully discrete cosmological $N$-body problem, derived from a linearization of the force from a perfect cubic periodic lattice\footnote{It is not limited to this case; the framework is equally valid for any Bravais lattice, such as body-centered and face-centered lattices.  We will focus on the simple cubic case, however.} using the \textit{dynamical matrix} formalism well known in solid state physics (see \S\ref{sec:PLT}).  As long as its perturbative assumption is satisfied, PLT fully describes the particle positions and velocities as a function of time (or redshift).  This allows analysis of discreteness effects by comparing the particle behavior for finite $N$ to the limit $N\rightarrow \infty$.

The authors of PLT have used their theory to quantify discreteness effects from the linear and weakly non-linear regimes \citep{Joyce_Marcos_2007b} to the fully non-linear regime \citep{Joyce+2009}.  However, PLT has not yet been used to improve the initial conditions of simulations.  In this work, we develop PLT-based corrections to the initial conditions that eliminate transients due to the initial grid configuration of the particles.  Additionally, we develop a fast and powerful new approach to second-order Lagrangian perturbation theory (2LPT) that does not rely on large Fourier transforms, and we demonstrate its accuracy by performing the actual particle evolution from $z=4999$ with our extremely precise $N$-body code \textsc{Abacus}.  We compare our answer to that of a well-known 2LPT code and find excellent agreement on all but small scales, where we expect differences due to the different assumptions inherent in our approaches.

Broadly speaking, simulations must produce power spectra and halo properties accurate to 1\% to support current and upcoming galaxy surveys \citep[e.g.][]{Tinker+2008, Weinberg+2013}.  Specifically, projects like the DES \citep{Frieman+2013}, LSST \citep{LSST_2012}, and \textit{Euclid} \citep{Cimatti+2009,Laureijs_2009} are projected to require 1\% accuracy in the matter power spectrum to $k=10\,h~{\rm Mpc}^{-1}$ \citep{Schneider+2016}.  These stringent demands are our motivation for careful examination and improvement of the initial conditions on small scales.

In \S\ref{sec:PLT}, we review the formalism of PLT.  Then, in \S\ref{sec:ICs}, we discuss our application of PLT to the initial conditions of cosmological simulations and quantify the improvements.  In \S\ref{sec:2lpt}, we develop our new approach to 2LPT and test its accuracy.  Finally, in \S\ref{sec:cosmo}, we discuss the implications of our findings for cosmological measurables derived from $N$-body simulations (halo masses, clustering, and power spectra), and summarize our results in \S\ref{sec:conc}.

\section{Particle Linear Theory}\label{sec:PLT}
Here, we review particle linear theory (PLT) as developed by \cite{Marcos+2006} \citep[see also][]{Joyce_Marcos_2007b,Joyce+2005}.  PLT gives the analytical evolution of a slightly perturbed lattice of self-gravitating particles, which is precisely the initial configuration of many cosmological simulations.  We will emphasize the ways in which this evolution diverges from that of the corresponding fluid system.

\subsection{PLT formalism}
Consider a simple cubic lattice of $N$ equal-mass particles in a box of side length $L$ with periodic boundary conditions.
In an expanding universe, the equation of motion is
\beq
{\ddot{\bfx}}_i + 2 H (t) {\dot{\bfx}}_i = -\frac{1}{a^3} \sum_{i\neq j} \frac{G m_j (\bfx_i-{\bf x}_j) }{|\bfx_i-\bfx_j|^3},
\label{eqn:eom}
\eeq
where $\bfx_i$ is the comoving position of particle $i$, $m_i$ is its mass, and $G$ is the universal gravitational constant.  $\bfx_i$ is related to the physical position by $\bfr_i = a(t)\bfx_i$, where $a(t)$ is the cosmological scale factor and $H(t) = \dot{a}/{a}$ is the corresponding Hubble factor.

If we label the original lattice site corresponding to particle $i$ with its comoving position $\bfR_i$, then we may write the displacement of particle $i$ from $\bfR_i$ as $\bfu(\bfR_i)$.  Thus, the full comoving position of a particle is given by $\bfx_i(t) = \bfR_i + \bfu(\bfR_i)$.  Following the convention in PLT, we generally drop the subscript $i$ from $\bfR_i$.

The right side of \re{eom} can be expanded at linear order in the relative displacements of particle pairs\footnote{When expanding the force in a Taylor series, one finds that each term in the sum over lattice sites $\bfR'$ has its own convergence criterion: $|\bfR - \bfR'| > |\bfu(\bfR) - \bfu(\bfR')|$.  This lends some robustness to the expansion, because even as particles move and some particle pairs start to violate this condition, many others may continue to satisfy it and thus still produce a useful approximation of the total force.} to yield
\beq
{\ddot \bfu}(\bfR,t) +2 H \bfu(\bfR,t) = -\frac{1}{a^3} \sum_{\bfR'} {\cal D} (\bfR - \bfR') \bfu(\bfR',t).
\label{eqn:linearized-eom}
\eeq
The matrix $\mathcal{D}(\bfR)$ is known in solid state physics as the \textit{dynamical matrix} \citep{Pines_1964}. For a given $\bfR$, $\mathcal{D}(\bfR)\bfu(\bfR)$ is the force induced at the origin by a particle at $\bfR$ displaced by $\bfu(\bfR)$.  Specifically, the dynamical matrix can be written as
\bea
{\cal D}_{\mu \nu} (\bfR \neq {\bf 0}) =Gm\left(\frac{\delta_{\mu \nu}}{R^3}-3\frac{R_\mu R_\nu}{R^5}\right) \\
{\cal D}_{\mu \nu} ({\bf 0}) = -\sum_{\bfR \neq {\bf 0}} {\cal D}_{\mu \nu} (\bfR),
\eea
where $\delta_{\mu \nu}$ is the Kronecker delta.  The second equation is a statement of Newton's third law.  $\mathcal{D}$ cannot be computed as simply as these expressions suggest, however, because there is an implicit sum over infinite periodic copies.  Ultimately, this means one must either compute $\mathcal{D}$ using an Ewald-type summation \citep[as in][]{Marcos+2006} or with a very precise $N$-body force solver, as we use (see \S\ref{sec:eigenmodes}).

In \re{linearized-eom}, $\mathcal{D}$ acts as a convolution kernel acting on the displacements, and thus it is not surprising that it has a natural action in Fourier space.
If we define the discrete Fourier transform and its inverse\footnote{See \cite{Marcos+2006} for subtleties regarding the summation limits.} as
\begin{align}
\tilde\bfu({\bfk},t) &= \sum_\bfR e^{-{\rm i} \bfk\cdot\bfR}\bfu(\bfR,t) \\
\bfu(\bfR,t) &= \frac{1}{N} \sum_\bfk e^{{\rm i} \bfk\cdot\bfR} \tilde\bfu(\bfk,t),
\label{eqn:discreteFT}
\end{align}
then we may write the equation of motion (\re{linearized-eom}) as
\beq
\ddot{\tilde\bfu}(\bfk,t) + 2 H (t) \dot{\tilde\bfu}(\bfk,t) = -\frac{1}{a^3} \tilde{\mathcal{D}}(\bfk) \tilde\bfu(\bfk,t).
\label{eqn:fourier-eom}
\eeq
We define $\tilde{\mathcal{D}}$ as the Fourier transform of $\mathcal{D}$, in analogy with \re{discreteFT}.  From the symmetry properties of $\mathcal{D}(\bfR)$, $\tilde{\mathcal{D}}(\bfk)$ must be a real, symmetric operator with three orthogonal eigenvectors $\bfe_n(\bfk)$ and eigenvalues $\omega^2_n(\bfk)$.

Because the eigenvectors of $\mathcal{D}$ form a complete basis at every $\bfk$, we can project an arbitrary displacement field onto the basis $\hat\bfe_n(\bfk)$.  Or, as we discuss in \S\ref{sec:spatial}, we can construct a displacement field that consists of one eigenmode at every $\bfk$.  For now, we will discuss the evolution of an arbitrary displacement field from initial conditions $\bfu(\bfR,t_0)$ and $\dot\bfu(\bfR,t_0)$.

We can represent the Fourier space evolution of $\tilde\bfu(\bfR,t)$ as a sum of the independent evolution of each eigenmode\footnote{Recall that modes at different wavevectors evolve independently in linear theory.}:
\begin{align}
\bfu(\bfk,t) = \sum_{n=1}^3 &U_n(\bfk,t)\left[\hat\bfe_n(\bfk)\cdot\tilde\bfu(\bfk,t_0)\right]\hat\bfe_n(\bfk) \nonumber\\
&+ V_n(\bfk,t)\left[\hat\bfe_n(\bfk)\cdot\dot{\tilde\bfu}(\bfk,t_0)\right]\hat\bfe_n(\bfk).
\label{eqn:fourier-evo}
\end{align}
The functions $U_n(\bfk,t)$ and $V_n(\bfk,t)$ can be found by substituting the above equation into \re{fourier-eom}, with the boundary conditions
\begin{align}
\nonumber
&U_n(\bfk,t_0)=1, &\dot{U}_n(\bfk,t_0)=0,\\ 
&V_n(\bfk,t_0)=0, &\dot{V}_n(\bfk,t_0)=1.
\label{eqn:normalization}
\end{align}
Replacing $\tilde{\mathcal{D}}\hat\bfe_n$ by $\omega^2_n\hat\bfe_n$, one finds
\beq
{\ddot{f}} + 2 H{\dot{f}}= -\frac{\omega_n^2({\bf k})}{a^3} f,
\label{mode-equation}
\eeq
to which $U_n(\bfk,t)$ and $V_n(\bfk,t)$ are the solutions.

The exact form of $U_n$ and $V_n$ depends on the cosmology.  In \LCDM{}, the early universe is well described by an Einstein-deSitter ($\Omega_m = 1$) cosmology, with scale factor $a(t) \propto t^{2/3}$ and Hubble constant $H(t) = 2/3t$.  Since the small-displacement (and thus early-time) regime is exactly what we are considering here, EdS is a good approximation to \LCDM{}.  Thus, we have
\begin{subequations}
\begin{align}
U_n(\bfk,t)=&\tilde\al(\bfk)\left[\al_n^{+}(\bfk)\left(\frac{t}{t_0}\right)^{\al_n^{-}(\bfk)}+\al_n^{-}(\bfk)\left(\frac{t}{t_0}\right)^{-\al_n^{+}(\bfk)}\right]\\
V_n(\bfk,t)=&\tilde\al(\bfk)t_0\left[\left(\frac{t}{t_0}\right)^{\al_n^{-}(\bfk)}-\left(\frac{t}{t_0}\right)^{-\al_n^{+}(\bfk)}\right]
\end{align}
\label{eqn:uv}
\end{subequations}
where
\beq
\tilde\al(\bfk)=\frac{1}{\al_n^{-}(\bfk)+\al_n^{+}(\bfk)}
\eeq
and
\begin{subequations}
\begin{align}
&\al_n^{-}(\bfk)=\frac{1}{6}\left[\sqrt{1+24\epsilon_n(\bfk)}-1\right]\\
&\al_n^{+}(\bfk)=\frac{1}{6}\left[\sqrt{1+24\epsilon_n(\bfk)}+1\right].
\end{align}
\label{eqn:alpha-epsilon}
\end{subequations}
In these expressions, $\epsilon_n(\bfk)$ are the normalized eigenvalues, given by
\beq
\epsilon_n (\bfk)\equiv -\frac{\omega_n^2(\bfk)}{4\pi G\rho_0}.
\label{eqn:def-epsilon}
\eeq

This completes the PLT description of the evolution of an arbitrary displacement and velocity field in an EdS universe, up to the numerical computation of the eigenmodes of $\mathcal{D}$ (see \S\ref{sec:eigenmodes}).

\subsection{Discreteness effects and the fluid limit}\label{sec:fluid}
We now have a quantitative framework in which to compare particle lattice evolution to the evolution of the equivalent fluid system.  Namely, we can compare the behavior of wave modes on the lattice to wave modes in the fluid system.  We discuss two ways in which discreteness manifests: deviation of the eigenvalue spectrum from unity, and deviation of the longitudinal eigenvectors from $\hat\bfk$.

\begin{figure}
\centering
\includegraphics[width=\linewidth]{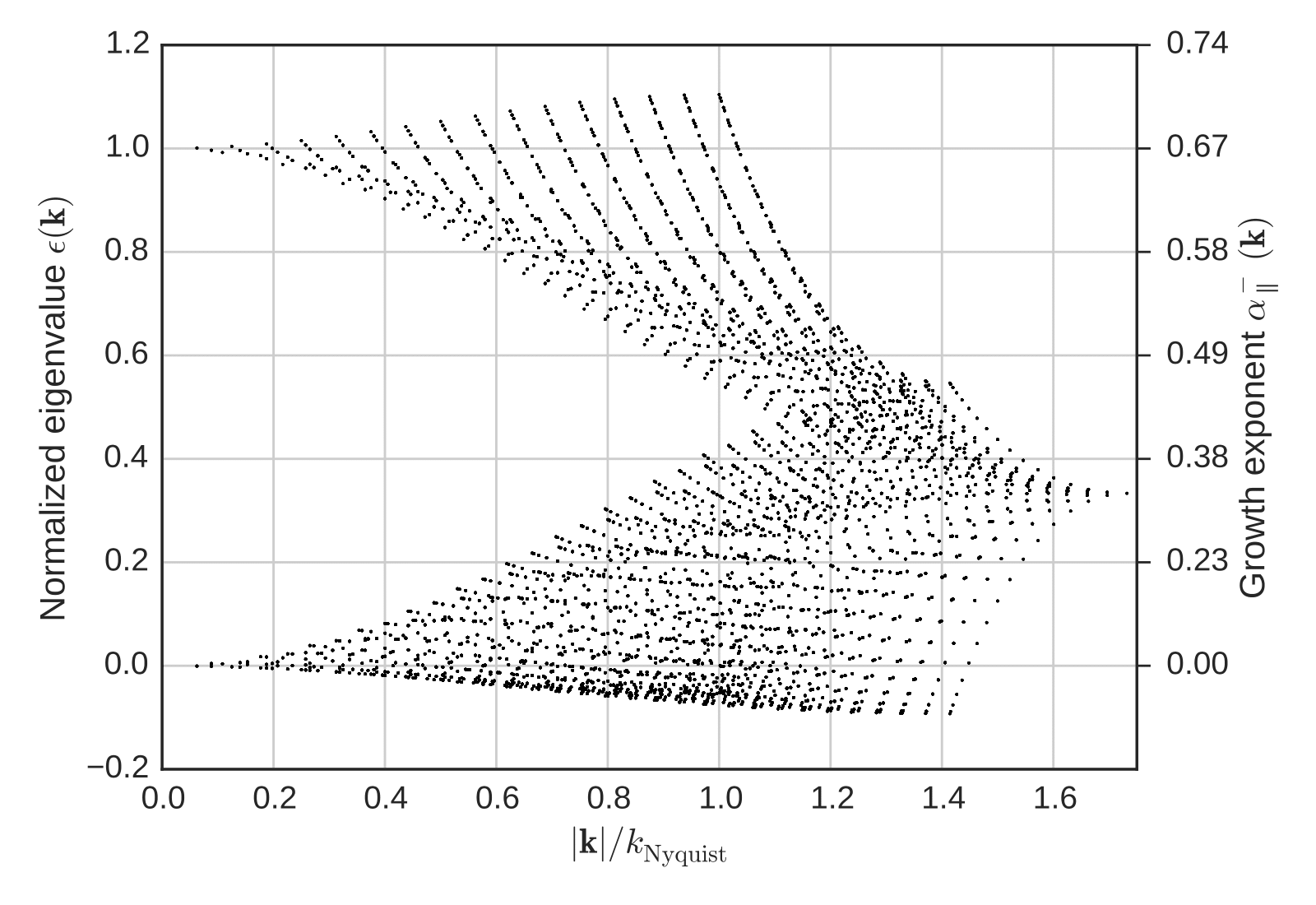}
\caption{Eigenvalue spectrum for a $32^3$ particle simple cubic lattice.  Eigenvalues of 1 and 0 correspond to fluid behavior for longitudinal and transverse modes, respectively (\re{def-epsilon}).  The corresponding growing mode exponent is labeled on the right axis (Eqs.~\ref{eqn:alpha-epsilon} \& \ref{eqn:simplified-fourier-evo}), where $2/3$ is the nominal fluid linear theory value.  Compare with \protect\cite{Joyce_Marcos_2007b} fig.~1.}
\label{fig:spectrum}
\end{figure}

In Fig.~\ref{fig:spectrum}, all three eigenvalues are plotted for every $\bfk$.  If the lattice behaved as a fluid, two of the eigenvalues would be 0 and one would be 1 at every $\bfk$.  The two null eigenvalues correspond to transverse modes, or modes with zero divergence and non-zero curl that do not source forces in fluid theory, so their deviation from 0 in Fig.~\ref{fig:spectrum} is purely an artefact of discreteness.  The eigenvalue of 1 corresponds to a longitudinal mode that produces density perturbations that source a force directly proportional to the overdensity.  The presence of $\epsilon_n > 1$ corresponds to an ``overdriven'' mode that collapses faster than the fluid limit, while $\epsilon_n < 1$ is an ``underdriven'' mode that collapses more slowly.  A mode with $-1/24 < \epsilon_n < 0$ is purely decaying; $\epsilon_n < -1/24$ is oscillatory.  Note that the eigenvalues converge to either 1 or 0 as $|\bfk| \rightarrow 0$, which is a reflection of the fact that we recover the fluid behavior in the limit of a well-sampled mode.

The orientation of modes explains why some modes are overdriven and some are underdriven for the same $|\bfk|$.  Modes aligned with the grid axes collapse faster than those skew to them.  This orientation dependence is a direct violation of isotropy.

\begin{figure}
\centering
\includegraphics[width=\linewidth]{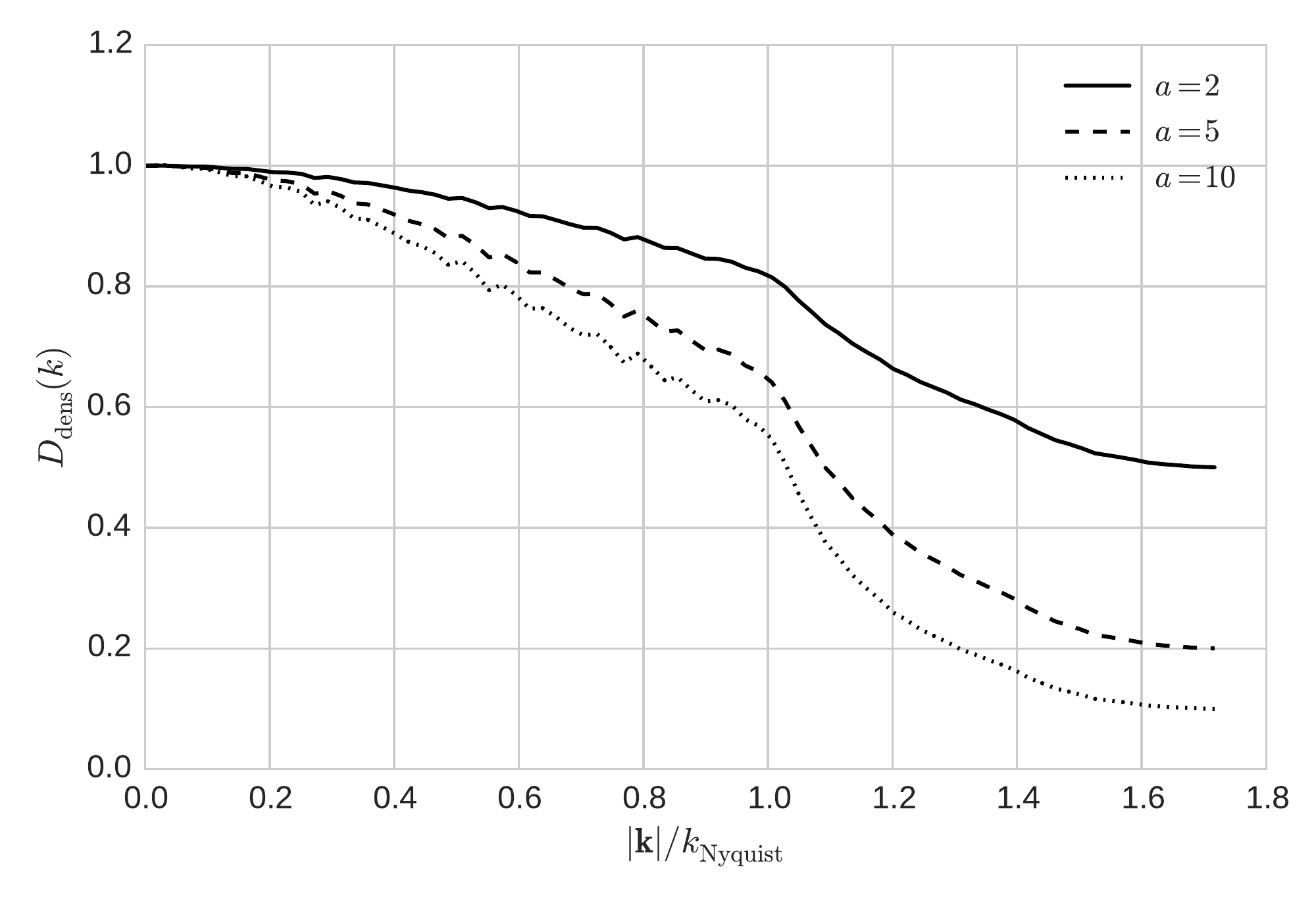}
\caption{Discreteness factor $D_{\rm dens}$ for a $64^3$ particle simple cubic lattice, averaged in bins of $|\bfk|$.  This gives the ratio of the density power spectrum in PLT to fluid theory as a function of scale factor (for $a_{\rm init} = 1$) and wavenumber.  Compare with \protect\cite{Joyce_Marcos_2007b} fig.~3, but note the differences due to our definition of $D_{\rm dens}$.}
\label{fig:Ddens}
\end{figure}

Due to the fact that some modes are consistently underdriven and some overdriven, we would expect accumulation of this effect over time.  In Fig.~\ref{fig:Ddens}, we plot the magnitude of this effect, averaged over mode orientations.  This plot illustrates one of the most surprising and important results of \cite{Marcos+2006}, which is that the power spectrum at a fixed redshift of a particle system \textit{diverges} from the fluid limit as $z_{\rm init} \rightarrow \infty$, because an earlier starting time means more time for these non-fluid effects to build up\footnote{This prediction of suppression of small-scale power with increasing initial redshift has been borne out in empirical tests, e.g.~\cite{LHuillier+2014}.}.  This divergence is particularly important to note because standard practice is to increase $z_{\rm init}$ and claim that the results are representative of the desired fluid behavior.  While this achieves the goal of reducing higher-order effects, the fact that the first-order results diverge from the fluid limit is often neglected.  Increasing the redshift as a test of convergence is only a safe procedure when considering scales much larger than the interparticle spacing.  The correct way, then, to test for convergence on intermediate and small scales is to increase the particle density while keeping the initial redshift fixed and compare the results at the same wavenumber.  This is the procedure we employ throughout this work.  Additionally, we attempt to make a correction for this effect by modifying the initial conditions, which we describe in \S\ref{sec:rescaling}.

The second manifestation of discreteness in PLT is in the eigenvectors of the dynamical matrix.  In fluid theory, only longitudinal (compressional) modes feel forces because they are the only modes that produce density contrasts.  This corresponds to the eigenmodes
\begin{align}
\hat\bfe_1 &= \hat\bfk; & \epsilon_1 &= 1; \nonumber\\
\hat\bfe_2 &= \hat\bfk_{2\perp}; & \epsilon_2 &=0; \nonumber\\
\hat\bfe_3 &= \hat\bfk_{3\perp}; & \epsilon_3 &= 0;
\label{eqn:fluid_modes}
\end{align}
where $\hat\bfk_{2\perp}$ and $\hat\bfk_{3\perp}$ are chosen such that $\{\hat\bfe_n\}$ forms an orthogonal basis.  In PLT, all three eigenvectors generically have non-zero eigenvalues and the longitudinal eigenvector $\hat\bfe_1 \ne \hat\bfk$.  Since simulations are nearly always initialized with purely $\hat\bfk$ modes, this manifests as forces misaligning with displacements.  This introduces vorticity that should eventually decay relative to the growing mode, but such effects do not disappear quickly, especially in higher-order statistics \citep{Scoccimarro_1998}.  See \S\ref{sec:spatial} for our correction of this effect.  The nearly-perfect alignment of forces and displacements that we achieve is also important for our implementation of second-order Lagrangian perturbation theory corrections (see \S\ref{sec:2lpt}).

\subsection{Numerical computation of $\mathcal{D}$}\label{sec:eigenmodes}
\re{fourier-evo} gives the analytical particle evolution in PLT but depends on knowing the eigenvectors $\hat\bfe_n(\bfk)$ and eigenvalues $\omega^2_n(\bfk)$ of $\tilde{\mathcal{D}}(\bfk)$, which must be calculated numerically.  \cite{Marcos+2006} compute the spectrum with a custom Ewald summation method, which works by decomposing the gravitational potential into near-field and far-field components.  The former is summed in configuration space and the latter in Fourier space, since they converge quickly in those respective spaces.  The sums are truncated when the series is determined to have converged.  The potential yields $\mathcal{D}(\bfR)$, which is then is Fourier transformed to $\tilde{\mathcal{D}}(\bfk)$.  Recall that $\tilde{\mathcal{D}}(\bfk)$ is a $3\times 3$ matrix at every $\bfk$, so the determination of the $3N$ eigenvectors and eigenvalues reduces to $N$ $3\times 3$ matrix diagonalizations, which can be done with any numerical linear algebra package.

Rather than build a custom Ewald summer, we take advantage of the high force accuracy of our $N$-body code \textsc{Abacus} (see \S\ref{sec:abacus}) and calculate $\mathcal{D}$ in the following manner:
\begin{enumerate}
\item Generate a uniform grid of $N$ particles;
\item Displace one particle by a small fraction of the interparticle spacing ($10^{-5}$ is sufficiently small) along the $x$-axis;
\item Measure the force induced on all other particles by this displaced particle and call this field $\bfF_{+x}(\bfR)$;
\item Displace the particle by the same amount in the $-x$ direction;
\item Measure the force and call this field $\bfF_{-x}(\bfR)$;
\item Add the forces to cancel second-order effects: $\frac{1}{2}(\bfF_{+x}(\bfR) - \bfF_{-x}(\bfR))/10^{-5}$ is one row of $\mathcal{D}(\bfR)$;
\item Repeat steps (ii) -- (vi) for the $y$- and $z$-axes.
\end{enumerate}
Having formed $\mathcal{D}(\bfR)$, we can now proceed exactly as before to calculate $\tilde{\mathcal{D}}(\bfk)$ and its eigenmodes.  In practice, we do not displace the particle along the $y$- and $z$-axes.  Instead, we permute the indices of the $x$ result to obtain the $y$ and $z$ results.  Furthermore, the eigenvalues and eigenvectors vary smoothly below $\kNy$, so rather than generate new eigenmodes for every different $N$, we generate one $N = 128^3$ grid and trilinearly interpolate to finer grids on-the-fly.

\subsection{\textsc{Abacus}: $N$-body cosmology to machine precision}\label{sec:abacus}
Throughout this work, we employ \textsc{Abacus}, an $N$-body code for cosmological simulations (Ferrer et al., in prep.; Metchnik \& Pinto, in prep.).  It is both extremely fast and accurate, capable of computing over 100 billion pairwise force interactions per second on a single computer node, with the option to compute forces to within nearly machine precision while maintaining competitive speeds.  It derives its performance from a combination of novel computational techniques and high-performance commodity hardware (GPUs and RAID disk arrays).

The computational domain in \textsc{Abacus} is divided into a grid of $\texttt{CPD}^3$ cells, where $\texttt{CPD}$ is the number of cells per dimension.  The force computation is split into near-field and far-field components based on this cell decomposition.  However, unlike most $N$-body methods, such as P\textsuperscript{3}M, this decomposition is exact -- the pairwise force between any two particles is always given exactly by either the near-field or far-field force.

Particles interact via the near-field force if they are within \texttt{NearFieldRadius} cells of one another.  For example, \texttt{NearFieldRadius} = 2 means a cell's 124 nearest neighbor cells are included in the near-force calculation.  We compute the force as a direct summation of $1/r^2$ forces (or some appropriately softened form).  This calculation can be accelerated with GPUs, which enables the impressive single-node performance.

Particles interact via the far-field force if they are separated by more than \texttt{NearFieldRadius} cells.  \textsc{Abacus} computes a multipole expansion of the particles in each cell using a variant of the fast multipole method (Metchnik \& Pinto, in prep.; see also \cite{Nitadori_2014} for a similar method).  This multipole expansion is then convolved with a derivatives tensor to yield a set of coefficients for a Taylor-series expansion of the force in a given cell.  The derivatives tensor is a fixed property of the grid for a given \texttt{CPD}, \texttt{NearFieldRadius}, and series expansion \texttt{Order}, and is thus precomputable.  The multipole computation, derivatives convolution, and evaluation of the Taylor series happens for each timestep.  Our performance tuning strategy is to change \texttt{CPD} to balance the cost of the near- and far-field computations.

In Table \ref{tbl:abacus_params}, we define two sets of parameters that we will refer to as ``normal precision'' and ``high precision'' throughout the rest of this work.  We use high precision for evaluation of the dynamical matrix in \S\ref{sec:eigenmodes} and most of the tests of the correctness of our methods in \S\ref{sec:ICs} \& \S\ref{sec:2lpt} (that is, the $64^3$ and $256^3$ particle simulations).  We note the exceptions as they occur, which are generally for simulations to low $z$.  In high precision, the maximum force error reaches nearly machine precision, but is more expensive to evaluate (and the lack of softening makes it unsuitable for low-$z$ applications).  In our $720^3$ and $1440^3$ cosmology simulations in \S\ref{sec:cosmo}, we use normal precision.  In computing the dynamical matrix, the high precision multipole \texttt{Order} and \texttt{Precision} are both necessary for an accurate determination of the eigenmodes.

\begin{table}
\caption{\textsc{Abacus} code options}
\begin{tightcenter}
\begin{tabular}{|l|r|r|}
\hline
Parameter & Normal precision & High precision\\
\hline
\texttt{SofteningLength} & 1/8 particle spacing  & 0 \\
Softening technique & Plummer & None \\
\texttt{NearFieldRadius}  & 2 & 3\\
Multipole \texttt{Order} & 8 & 16\\
\texttt{Precision} & 32-bit & 64-bit \\
\hline
Max force error & $1 \times 10^{-4}$ & $2 \times 10^{-8}$\\
Median force error & $2 \times 10^{-6}$ & $4 \times 10^{-11}$\\
\hline
\end{tabular}
\end{tightcenter}
{\sc Notes} --
The force error is the maximum fractional error on a set of $2^{16}$ uniformly random distributed particles, compared to the true $1/r^2$ forces computed with an Ewald summation in 256-bit precision.  The other parameters are described in \S\ref{sec:abacus}.
\label{tbl:abacus_params}
\end{table}

The exceptional force accuracy of \textsc{Abacus} enables us to carry out the precise testing of the initial conditions in the following sections.

\section{Corrections to Initial Conditions}\label{sec:ICs}
In this section and the next, we discuss four applications of the above theory to improve the initial conditions of cosmological simulations:
\begin{enumerate*}[label=(\roman*)]
\item initializing every mode in the simulation to a single eigenmode at every $\bfk$;
\item correcting the velocities of every mode to eliminate decaying-mode transients;
\item rescaling the initial displacement amplitudes such that the power spectrum will match the linear prediction at a later time; and
\item calculating second-order Lagrangian perturbation theory corrections using a novel in-place scheme.
\end{enumerate*}
The last correction we defer to the next section, as its derivation is independent of PLT.

\subsection{Spatial transients}\label{sec:spatial}
As we know from our consideration of the eigenmodes of $\tilde{\mathcal{D}}$, every $\bfk$ has a three orthogonal eigenvectors $\hat\bfe_n$: one ``longitudinal'' and two ``transverse'' eigenvectors.  The longitudinal eigenvector is most closely parallel to $\hat\bfk$ (thus we label it $\hat\bfe_\parallel$), and it converges to $\hat\bfk$ as $|\bfk| \rightarrow 0$.  Below the Nyquist frequency, the longitudinal eigenmode also always has the largest eigenvalue, meaning it is the strongest growing mode on the grid.  This has the following implication. Consider a mode $\tilde\bfu(\bfk)$ oriented along $\hat\bfk$.  Generically, this mode will have non-zero components along all three PLT eigenvectors.  No matter the relative magnitudes of these components, the one with the largest eigenvalue will dominate after some time, because of the power-law behavior of \re{uv}.  Until then, the excitation of the transverse eigenmodes can be seen as a transient that is purely dependent on the time since initialization.  This is a discreteness effect that we can eliminate by initializing each mode in the longitudinal eigenmode $\hat\bfe_\parallel$ instead of $\hat\bfk$.  We will call this mode $\tilde\bfu_\parallel(\bfk)$, since $\tilde\bfu_\parallel(\bfk) \propto \hat\bfe_\parallel$.

What amplitude do we choose for $\tilde\bfu_\parallel(\bfk)$?  There are two reasonable choices:
\begin{align}
|\tilde\bfu_\parallel(\bfk)| = |\tilde\bfu(\bfk)| \quad{\rm or}\quad |\tilde\bfu_\parallel(\bfk)| = \frac{|\tilde\bfu(\bfk)|}{\hat\bfe_\parallel \cdot \hat\bfk}.
\label{eqn:gauge}
\end{align}
The former is simply a rotation of the old mode into the new direction, while the latter preserves the projection of the new mode onto $\hat\bfk$.  We choose the latter gauge because it preserves the density power spectrum.  In other words, the divergence remains unchanged, but we add a small curl component.

As always, we only excite modes below the Nyquist wavenumber, defined as $\kNy = \pi/\Delta x$, where $\Delta x$ is the particle spacing.  This is the maximum wavenumber at which one can inject power without aliasing to lower wavenumbers.  This corresponds to modes that are sampled by at least two particles per cycle.

\subsection{Temporal transients}\label{sec:temporal}
The above prescription guarantees that the displacements start in the longitudinal eigenmode of the grid; now we must turn to the growing mode.  This corresponds to choosing the initial velocities such that the decaying terms in Eqs.~\ref{eqn:uv} cancel each other when substituted into \re{fourier-evo}.  For an initial displacement field $\tilde\bfu_\parallel(\bfk,t_0)$, the velocity field that cancels the decaying terms is
\beq
\tilde\bfv_\parallel(\bfk,t_0) = \frac{\alpha^-(\bfk)\tilde\bfu_\parallel(\bfk,t_0)}{t_0},
\label{eqn:PLTvel}
\eeq
which, in combination with our choice above to use only $\hat\bfe_\parallel$, causes \re{fourier-evo} to simplify to
\beq
\bfu(\bfk,t) = \left(\frac{t}{t_0}\right)^{\alpha^-_\parallel (\bfk)}\bfu(\bfk,t_0).
\label{eqn:simplified-fourier-evo}
\eeq
In other words, the displacements evolve in the pure growing mode.

Note that this velocity choice is a significant departure from the Zel'dovich approximation \citep{Zeldovich_1970} in which the velocities are always parallel to the displacements.  In our prescription, the parallel property still holds true in Fourier space, but the $\bfk$ dependence of $\alpha^-$ means it will not hold in configuration space.

We have now shown how to establish displacement and velocity fields $\bfu_\parallel(t_0)$ and $\bfv_\parallel(t_0)$ for an arbitrary input power spectrum that will eliminate transients to linear order.  Two issues remain: the systematic under-/over-growth of modes on the grid, and non-linear transients.  We first turn to the former.

\subsection{Growth rates and rescaling}\label{sec:rescaling}
As we discussed in \S\ref{sec:fluid}, every eigenvalue in Fig.~\ref{fig:spectrum} not equal to 1 will grow faster or slower than fluid theory predicts.  The effects can be significant: the average undergrowth of a mode at $k_{\rm Nyquist}/2$ is about $15\%$ after a factor of 10 increase in scale factor (see Fig.~\ref{fig:Ddens}).  To achieve $1\%$ accuracy in the power spectrum, we would have to limit ourselves to wavenumbers below $\sim k_{\rm Nyquist}/10$ (if the problem were purely linear).  Systematic small-scale undergrowth could also impact non-linear clustering, which we investigate in \S\ref{sec:cosmo}.  Furthermore, the growth rates are dependent on the orientation relative to the grid, which could imprint preferred axes on the clustering.

Can we correct for this effect?  Fluid linear theory gives us the expected displacement power spectrum as a function of time, and \re{fourier-evo} gives the \textit{actual} power spectrum that will be produced in a simulation.  Thus, we can try rescaling the initial power by the ratio\footnote{In practice, we are rescaling displacement amplitudes, not power, so we rescale by $\sqrt{D_{\rm dens}(\bfk, t)}$.}
\begin{align}
D_{\rm dens}(\bfk, t) &\equiv \frac{P^{\rm PLT}(\bfk,t)}{P^{\rm fluid}(\bfk,t)} = \frac{|\tilde\bfu^{\rm PLT}(\bfk,t)\cdot \hat\bfk|^2}{|\tilde\bfu^{\rm fluid}(\bfk,t)\cdot \hat\bfk|^2} \nonumber \\
&=  \frac{\left|\left(\frac{t}{t_0}\right)^{\alpha^-_\parallel (\bfk)} \frac{\tilde u^{\rm fluid}(\bfk,t_0) \hat\bfe}{\hat\bfe\cdot \bfk}\cdot \hat\bfk\right|^2}{|\tilde\bfu^{\rm fluid}(\bfk,t)\cdot \hat\bfk|^2} \nonumber \\
&= \left(\frac{a(t)^{3\alpha^-_\parallel(\bfk)/2}}{a(t)}\right)^2,
\end{align}
where in the second line we have used Eqs.~\ref{eqn:gauge} \& \ref{eqn:simplified-fourier-evo}.  This requires selecting a ``target redshift'' $z_{\rm target}$ at which time the simulation power spectrum and fluid power spectrum will match in linear theory.  This redshift should be early enough that the displacements are still perturbative and PLT is still valid, but close enough to the onset of non-linear evolution that the non-linearities will be seeded with the correct power spectrum.  In our tests in \S\ref{sec:cosmo}, we try $z_{\rm target} = 12$ and $5$.  One still expects these ``growth rate'' effects to be present during non-linear evolution, but PLT loses descriptive power in that regime, so we can no longer apply rescaling.

\begin{figure}
\centering
\includegraphics[width=\linewidth]{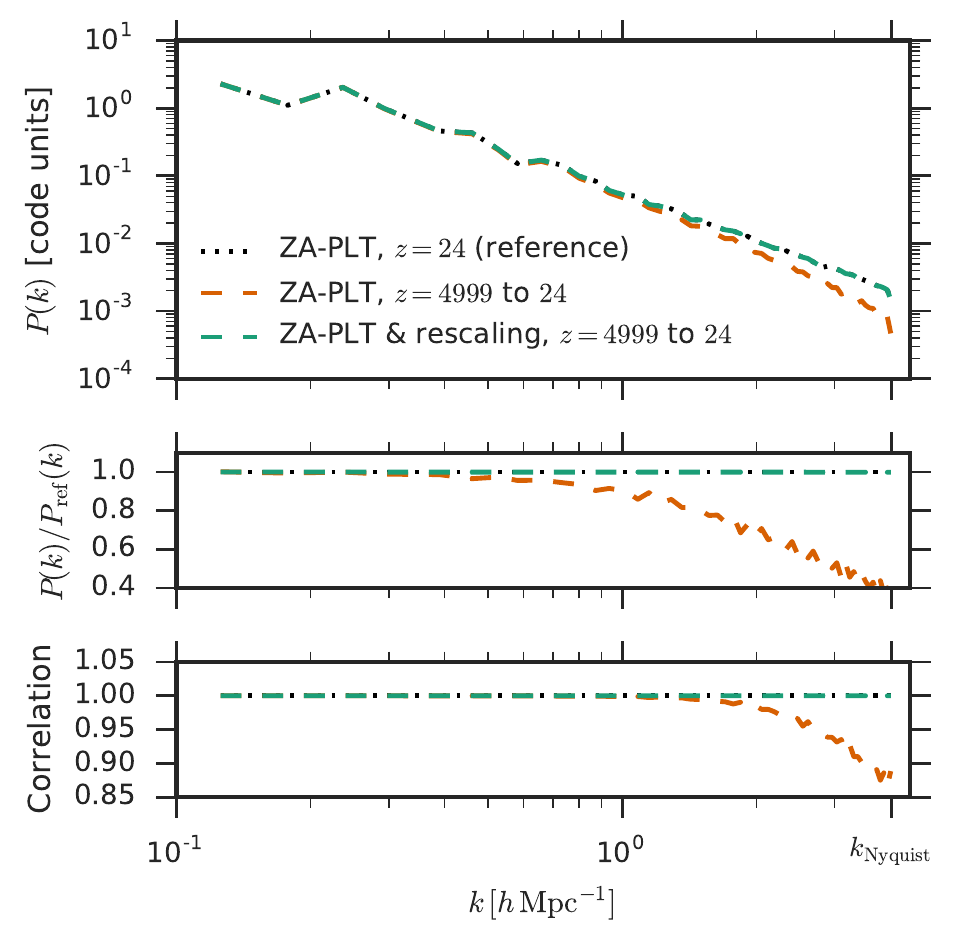}
\caption{A test of rescaling in the strongly linear regime (evolving a simulation from $z=4999$ to $z=24$ and using $\sigma_8 = 0.8/1000$ at $z=0$).  \textit{Top}: The density power spectrum for the reference (input) power spectrum at $z=24$ (black dotted line), and the simulation power spectra at $z=24$ with and without rescaling (dashed lines).  The green dashed line is not the reference theory line -- it is a simulation output -- but it matches the reference to a factor of $10^{-5}$ in the displacements.  This demonstrates that rescaling is capable of completely restoring the predicted fluid power spectrum in the strongly linear regime.  \textit{Middle}: The ratio of the simulation power spectra with the reference.  \textit{Bottom}: The cross-correlation of the phases of the  density fields with the reference (see Fig.~\ref{fig:2LPT_accuracy} caption for details).
}
\label{fig:rescaling}
\end{figure}

One major concern (and indeed the concern that \cite{Joyce_Marcos_2007b} raise) with this ``rescaling'' is the accumulation of non-linearities while evolving from $z_{\rm init}$ to $z_{\rm target}$ due to the larger self-interaction of the displacements, since the displacement field is offset from the ``true'' fluid value during this time.  In the strongly linear regime, this is a demonstrably negligible effect.  To quantify this, we ran \textsc{Abacus} in a high-precision mode (see Table \ref{tbl:abacus_params}) from $z_{\rm init} = 4999$ to $z_{\rm final} = 24$, a factor of 200 in scale factor, for $64^3$ particles in a $50\,h^{-1}$Mpc box.  Furthermore, we decreased $\sigma_8$ (the normalization of the power spectrum) by a factor of 1000 to decrease the displacement amplitudes, such that PLT should fully describe the evolution of the system.  We tested two initial conditions: one with rescaling and one without (both started from the Zel'dovich approximation in the PLT growing mode) and compare both to the linear theory prediction at $z = 24$.  The results are shown in Fig.~\ref{fig:rescaling}, which demonstrates the remarkable success of rescaling.  In configuration space, the particle displacements and velocities match linear theory to 0.006\% on average\footnote{Our definition of particle-averaged fractional error is the average absolute error over the mean magnitude, or $\langle|{\bf a}-{\bf b}|\rangle/\langle|{\bf a}+{\bf b}|/2\rangle$  We adopt this definition to avoid large numerical scatter due to the tiny magnitudes of some of the displacements.},
versus 6\% without rescaling.  In Fourier space, rescaling fully restores a 60\% power deficit at $\kNy$.  Besides being a strong confirmation of the correctness of PLT and rescaling within their regime of applicability, this is a remarkable testament to \textsc{Abacus}'s ability to evolve a system with displacements of order $10^{-6}$ of the interparticle spacing.

Of greater interest is rescaling in the weakly non-linear regime, where we intend to apply it in practice (for example, starting a simulation at $z_{\rm init} = 49$ with $z_{\rm target} = 5$).  We anticipate that non-linearities may arise from two sources: fluid non-linearities that are present in the true physical problem and non-physical non-linearities due to the offset evolution of the rescaled field before $z_{\rm target}$.  Thus, we must test whether the latter are sufficiently small.  To that end, we run a simulation identical to the above, but with $z_{\rm init} = 49$ and $z_{\rm target} = 5$ and 2LPT initial conditions (see \S\ref{sec:2lpt}), and compare it to a simulation oversampled by a factor of 4.  Specifically, we increase the particle count to $256^3$ and truncate the power spectrum at $k_{\rm Nyquist}/4$.  By only adding power below this wavenumber, we are oversampling the existing modes in the $64^3$ box which thus reduces the requisite amount of rescaling on those modes.  Thus, we expect non-linearities in the $256^3$ simulation to represent fluid non-linearities, not rescaling non-linearities.\footnote{To make our comparisons meaningful, we compare the density fields as computed by the divergence of the displacements fields.  In other words, we discard any curl components in the displacement fields.  This is because the PLT growing mode introduces a small curl component, which is smaller for the $256^3$ modes than the $64^3$ modes, but it is ultimately a fixed property of the grid and does not change the density power spectrum.}

\begin{figure}
\centering
\includegraphics[width=\linewidth]{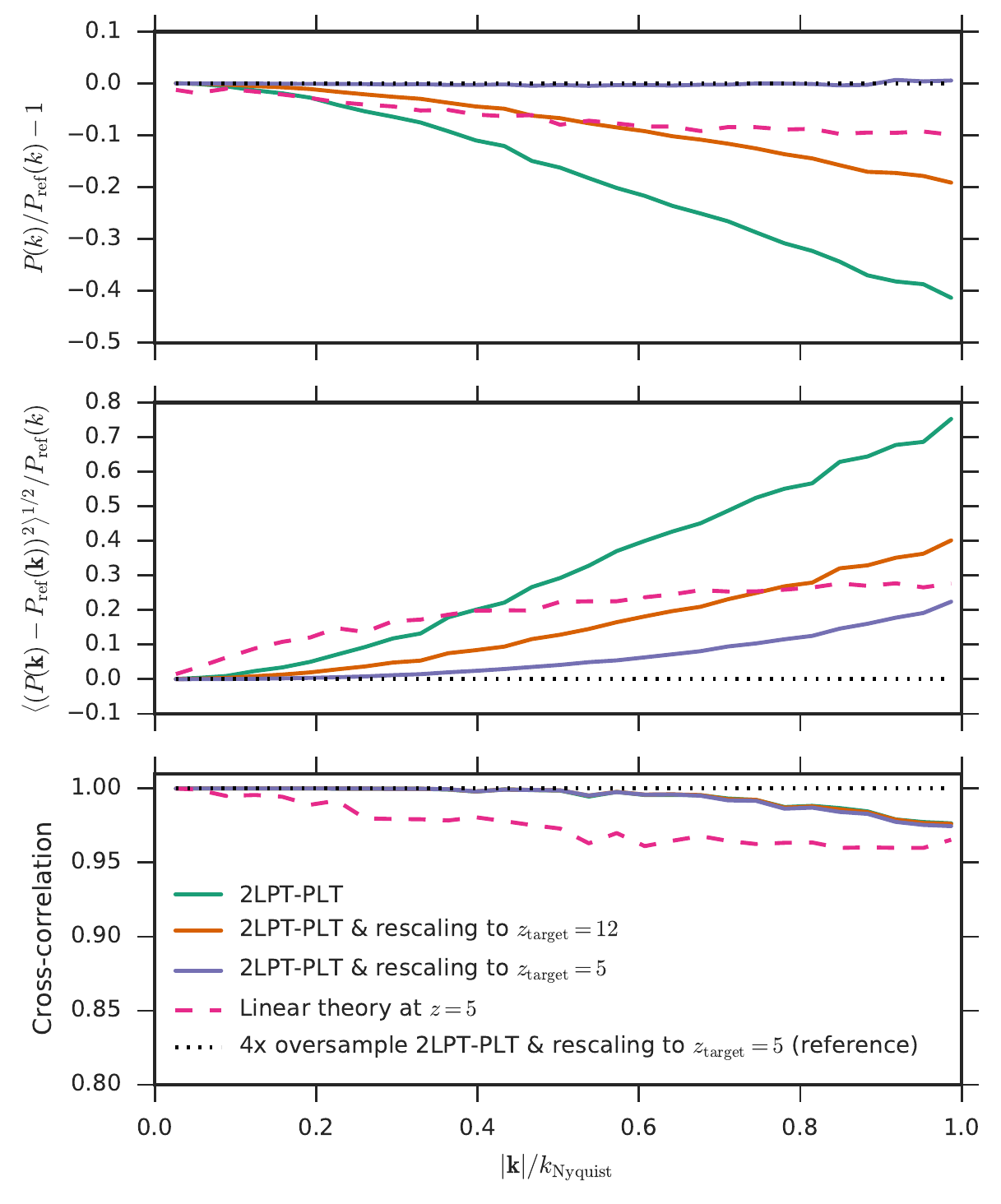}
\caption{A test of rescaling of the initial conditions as we intend to use it in practice (see \S\ref{sec:rescaling}). The cosmology is \LCDM\ with $\kNy = 4\,h~{\rm Mpc}^{-1}$ and a spline softening radius of $1/20$ of the particle spacing.  The reference is a $256^3$ particle oversampled simulation (black dotted line), and the three test cases are $64^3$ simulations with no rescaling, rescaling to $z_{\rm target} = 12$, and rescaling to $z_{\rm target} = 5$, respectively.  All are initialized at $z=49$ with 2LPT in the PLT growing mode.  The dashed line is the linear theory prediction for $z=5$, so the difference with the reference line quantifies the amplitude of non-linearities.  \textit{Top}: the ratio of each power spectrum with the reference.  Rescaling completely restores the power spectrum across the whole range of $k$.  \textit{Middle}: The root-mean-square deviation of $P(\bfk) = \tilde\delta(\bfk)\tilde\delta^*(\bfk)$ in bins of $k$.  See Fig.~\ref{fig:2LPT_accuracy} caption for details.  \textit{Bottom}: The cross-correlation of the phases of each density field with the reference simulation.
}
\label{fig:rescaling-NL}
\end{figure}

The results of this test are shown in Fig.~\ref{fig:rescaling-NL}, where we have also shown the results of $z_{\rm target} = 12$.  Rescaling to $z_{\rm target} = 5$ completely restores the lost power at all scales (up to 40\% at $\kNy$).  If there were substantial non-fluid non-linear effects, we would have expected excess power at $\kNy$ due to the earlier onset of non-linear growth, which is not present.  This is not a purely linear regime, either: the non-linear contribution to the power is about 10\% at $z=5$ (dashed line).  The scatter of $P(\bfk)$ about the reference (reaching 15\% at $\kNy$) may be evidence for these effects, but this is relatively unconcerning given that the uncorrected power spectrum has a 40\% power deficit at the same scale.  Thus, we consider $z_{\rm target} = 5$ a safe choice for use in cosmological simulations.  Rescaling does not substantially change the cross correlation, which is already very good on all scales.  This is consistent with our expectation from PLT that only the mode amplitudes are wrong relative to fluid linear theory; the phases remain unaffected.

We use 2LPT in this test because it is important for removing non-linear transients (as we show in the next section), but we choose a relatively high starting redshift of 49 to decrease its relative amplitude, since we want our results to be a measurement of rescaling, not 2LPT.

It should be emphasized that rescaling is only possible because of the eigenmode and growing mode corrections that we have already made.  Otherwise, we would not know the growth rate for any mode and could not compensate for it, as each would be a mixture of three growth exponents (some negative!).  In other words, the position corrections start the displacements in the longitudinal grid eigenmodes, the velocity corrections select the growing solution, and rescaling ``divides out'' the under-/overgrowth.

\section{Second-order Lagrangian Perturbation Theory in Configuration Space}\label{sec:2lpt}
We present next a new technique for computing non-linear displacement and velocity corrections with second-order Lagrangian perturbation theory (2LPT).  We emphasize that this scheme is derived from continuous theory -- not PLT -- so does not explicitly account for grid effects. However, the configuration-space approach we employ naturally preserves the displacements in the PLT longitudinal eigenmodes.  We numerically demonstrate the accuracy of our approach by comparing it with the actual evolution from very high redshift -- an approach only possible with PLT corrections.  We first present a derivation of our technique from continuous theory, then detail its implementation in \textsc{Abacus} and tests of its accuracy.

\subsection{Theory}\label{sec:2lpt_theory}
As before, we take $\bfx = \bfR + \bfu$ to be the comoving position, where $\bfR$ is the initial grid and $\bfu$ is the comoving Lagrangian displacement.  Taking $\bfv$ and $\bfg$ to be the comoving velocity and gravitational force, respectively, the equations of motion in an expanding universe are
\begin{gather}
\frac{d\bfx}{dt} = \bfv \\ \label{eqn:fluid_v_def}
\frac{d\bfv}{dt} + 2H\bfv = \bfg \\ \label{eqn:fluid_eom}
\nabla_x \cdot \bfg = -4\pi G \rho_{\rm comoving} a^{-3} \delta
\end{gather}
In linear theory, $\bfg = (3/2)\Omega_m H^2 \bfu$.
Substituting this and \re{fluid_v_def} into \re{fluid_eom} yields the equation of motion
\beq
\frac{d^2 \bfu}{dt^2} + 2H \frac{d \bfu}{dt} = \frac{3\Omega_m H^2}{2} \bfu.
\eeq
For $\Omega_m = 1$, we have $a \propto t^{2/3}$ and $H = 2/3t$, which gives the usual growing-mode solution $\bfu \propto t^{2/3} \propto a$.

Beyond linear theory, the relation $\bfu \propto \bfg$ will break down.  Thus, we can consider perturbative corrections to this field by writing
\beq
\bfu(t) = \epsilon d_1(t)\bfu_1 + \epsilon^2 d_2(t)\bfu_2 + \epsilon^3 d_3(t)\bfu_3 + O(\epsilon^4)
\eeq
where $\epsilon$ is a bookkeeping notation that represents the order of the term in this perturbative expansion.  The functions $d_n(t)$ and fields $\bfu_n$ are arbitrary at this point, but we will find them shortly by considering the dynamics.

If we take just the first term of the expansion ($\bfu(t) = \epsilon d_1(t)\bfu_1$) but also consider the interaction of the field with itself, then we must have a force series of the form
\beq
\bfg = \frac{3\Omega_m H^2}{2} [\epsilon d_1(t)\bfu_1 + \epsilon^2 d_1^2(t)\bfS(\bfu_1) + O(\epsilon^3)],
\label{eqn:O1_force}
\eeq
where $\bfS$ is some field that is second-order in $\bfu_1$.

Taking the next term in the expansion, we have $\bfu(t) = \epsilon d_1(t)\bfu_1 + \epsilon^2 d_2(t)\bfu_2$.  We must now consider the self-interaction of the first-order part, the cross-interaction of the first- and second-order parts, and the self-interaction of the second-order part.  This yields
\beq
\bfg = \frac{3\Omega_m H^2}{2} [\epsilon d_1(t)\bfu_1 + \epsilon^2 d_2(t)\bfu_2 + \epsilon^2 d_1^2(t)\bfS(\bfu_1) + O(\epsilon^3)],
\label{eqn:O2_force}
\eeq
where we have dropped any terms smaller than $\epsilon^2$.

Substituting this into the equation of motion, we have
\begin{multline}
\left(\frac{d^2}{dt^2} + 2H\frac{d}{dt}\right)(\epsilon d_1(t)\bfu_1 + \epsilon^2 d_2(t)\bfu_2) \\
= \frac{3\Omega_m H^2}{2}(\epsilon d_1(t)\bfu_1 + \epsilon^2 d_2(t)\bfu_2 + \epsilon^2 d_1^2(t)\bfS(\bfu_1)).
\end{multline}
Separating by order, we recover the linear growth equation
\beq
\left(\frac{d^2}{dt^2} + 2H\frac{d}{dt} - \frac{3\Omega_m H^2}{2}\right)d_1(t)\bfu_1 = 0,
\eeq
which has the growing-mode solution $d_1(t) \propto t^{2/3}$.  The next order is
\beq
\left(\frac{d^2}{dt^2} + 2H\frac{d}{dt} - \frac{3\Omega_m H^2}{2}\right)d_2(t)\bfu_2 = \frac{3\Omega_m H^2}{2}d_1^2(t)\bfS(\bfu_1).
\eeq
Thus, we find that $\bfS(\bfu_1) = \bfu_2$.  That is, as the linear displacement field grows, the first non-linear correction to the displacements is given by the part of the force due to interaction of the linear part with itself, up to an overall scaling.  The time dependence is simply given by solving the above ODE for $d_2(t)$, which yields $d_2(t) = (3/7) d_1^2(t)$.

How do we find $\bfS(\bfu_1)$?  If we write the force from $d_1(t)\bfu_1$ as $\bfg[d_1(t)\bfu_1]$, then \re{O1_force} tells us
\begin{align}
d_2(t)\bfu_2 &= \frac{3}{7}d_1^2(t)\bfS(\bfu_1) \nonumber\\
&= \frac{3}{7}\frac{2}{3\Omega_m H^2}\frac{1}{2}(\bfg[d_1(t)\bfu_1] + \bfg[-d_1(t)\bfu_1]).
\end{align}
Specifically, this is possible because $\bfS(\bfu_1)$ has even parity with respect to $\bfu_1$.  Furthermore, note that the $O(\epsilon^3)$ terms have canceled due to their odd parity.  In summary, two force calculations with opposing first-order displacements isolates the second-order displacements to third-order accuracy.

We can calculate the velocities at each order simply from $\bfv = \dot \bfu$.  For $\bfu = \sum d_j(t)\bfu_j$, 
\beq
\frac{d(d_j)}{dt} = d_j\frac{1}{d_j}\frac{d(d_j)}{da}\frac{da}{dt} = d_j H f_j,
\eeq
where $f_j$ is the familiar $d \ln{d_j}/d\ln{a}$, which is just a property of the cosmology; for example, with $\Omega_m = 1$, we have $d_1 \propto a$ and $d_2 \propto a^2$, so $f_1 = 1$ and $f_2 = 2$.  This yields velocities $\bfv = \sum H(t) f_j(t)d_j(t)\bfu_j$.

Having described our theory, we can now identify how it will interact with PLT and thus the particle grid.  Since our first-order displacements will be in the longitudinal eigenmode, they can only produce second-order forces also in the longitudinal eigenmode, in analogy with fluid $\hat\bfk$ modes being unable to produce forces with a curl.  Thus, the particle displacements after 2LPT will still be in the longitudinal eigenmode.  However, our fluid theory assumes that the second-order force is proportional to $a^2$ and that its direction is constant, both of which are not true in PLT.  As a displacement grows in PLT, the wavevector-dependent linear growth factors will cause it to change direction in configuration space, causing the force direction to change as well.  Thus, we attach the wrong displacement amplitudes and velocities.  Of course, we would have to Fourier transform the displacements to correct these effects, which would negate much of the advantage of our configuration space approach.  On large scales, we expect our corrections to be accurate, as the grid converges to fluid behavior.  This is the behavior we find in \S\ref{sec:2lpt_results}.
 
 \subsection{Implementation}\label{sec:2lpt_imp}
We implement our 2LPT as follows in \textsc{Abacus}.  Normally, every timestep contains a velocity update (``Kick'') and position update (``Drift'') for every particle; in the following, we express our 2LPT approach as a set of Kick and Drift operators.
\begin{enumerate}
\item Generate a field of first-order displacements $d_1(t)\bfu$ using any standard technique [like the Zel'dovich approximation (ZA)], preferrably with PLT corrections.  Apply the displacements to the particles.
\item Compute the force $\bfg[d_1(t)\bfu]$.  Store in the velocity.  This is like a Kick.
\item Reverse the displacement of every particle; that is, give every particle the position $\bfx = \bfR - d_1(t)\bfu$.  This requires retrieving the initial grid location $\bfR$, which we store in each particle's ID number.  This is like a Drift.
\item Compute the force $\bfg[-d_1(t)\bfu]$.  Add to the velocity.  This is like a Kick.
\item Take the position (currently holding the displacement $-d_1(t)\bfu$) and the velocity (currently holding $7 H^2\Omega_m d_2(t)\bfu_2$) and manipulate to form the second-order position $\bfR + d_1(t)\bfu_1 + d_2(t)\bfu_2$ and the second-order velocity $H d_1(t)\bfu_1 + 2 H d_2(t)\bfu_2$.  Store in the position and velocity.  This is like a Drift.
\end{enumerate}
The result is second-order Lagrangian perturbation theory for the cost of two force evaluations and memory requirements equal to the normal simulation code.

\subsection{Accuracy}\label{sec:2lpt_results}
\begin{figure*}
\centering
\includegraphics[width=\linewidth]{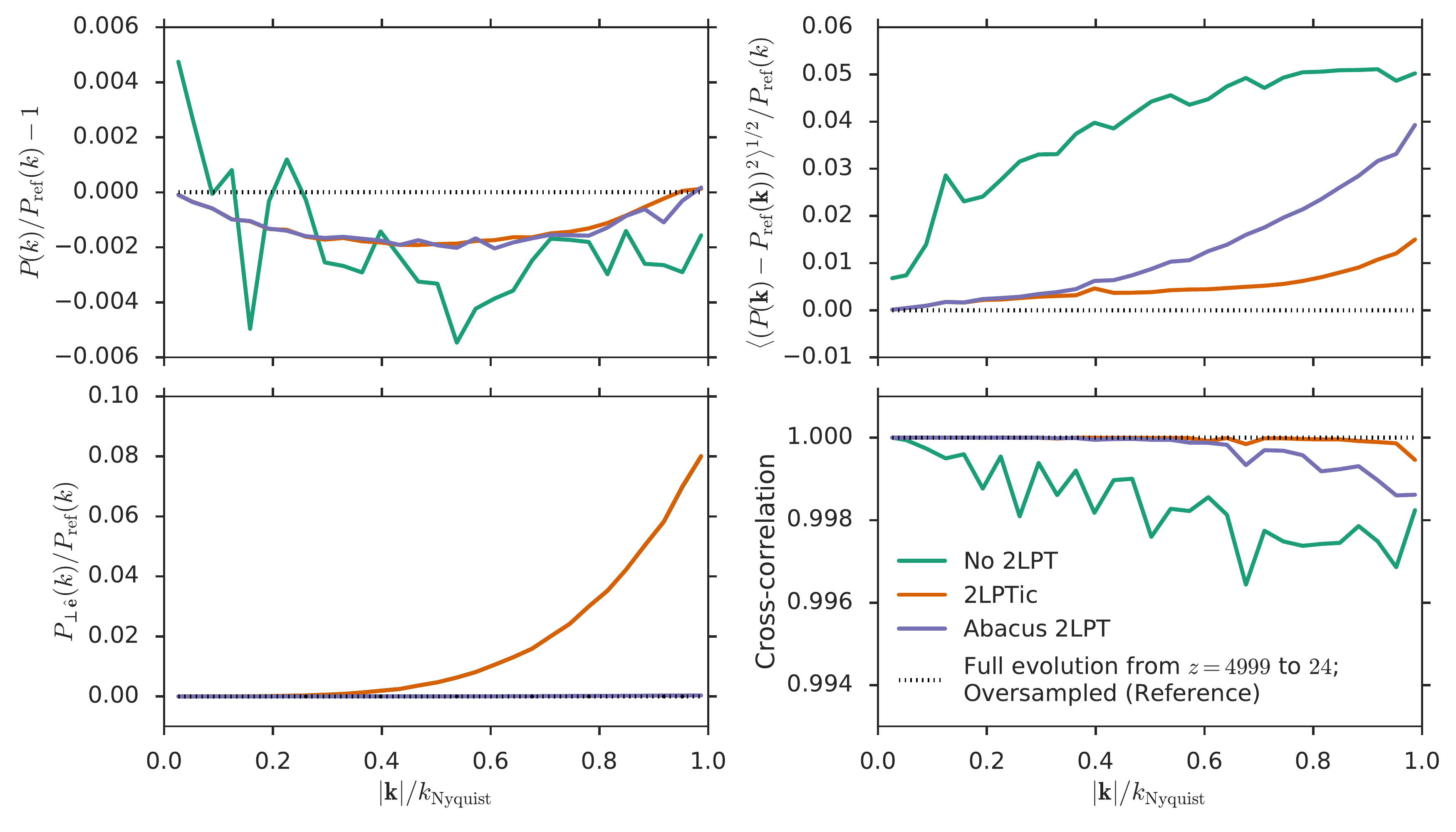}
\caption{A comparison of our configuration-space 2LPT (``\textsc{Abacus} 2LPT'') with a standard Fourier-space 2LPT code (``\textsc{2LPTic}'') at $z=24$ on a $64^3$ particle grid.  The reference solution is taken to be the full evolution from $z=4999$ of an oversampled, rescaled simulation in the PLT growing mode with $512^3$ particles as described in \S\ref{sec:2lpt_results}.  \textsc{Abacus} 2LPT has larger RMS deviations in $P(\bfk)$ versus \textsc{2LPTic} (top right; 4\% vs 1.5\% at $\kNy$), but no power $P_\perp(k)$ in transient curl modes (0\% versus 8\% at $\kNy$; bottom left). The mean power $P(k)$ and phase cross-correlation are excellent at all $k$ for both \textsc{Abacus} 2LPT and \textsc{2LPTic}.  \textit{Top left}: The ratio of the power spectrum $P(k)$ of each of the density fields with the reference.  We compute $P(k)$ from the density modes $\tilde\delta(\bfk)$ as $P(k) = \left<\tilde\delta(\bfk)\tilde\delta^*(\bfk)\right>$, where $^*$ denotes complex conjugation and $\left<\cdot\right>$ denotes averaging in annular bins of $k$.  We compute $\tilde\delta(\bfk)$ from the displacements $\tilde\bfu(\bfk)$ in Fourier space as $\tilde\delta(\bfk) = \bfk\cdot\tilde\bfu$. \textit{Top right}: The root-mean-square deviation of $P(\bfk) = \tilde\delta(\bfk)\tilde\delta^*(\bfk)$ in bins of $k$.  \textit{Bottom left}: Transverse (curl-mode) power, measured relative to the longitudinal PLT eigenvector $\hat\bfe_\parallel(\bfk)$.  The transverse power is computed as $P_{\perp\hat\bfe}(k) = \left<\tilde\delta_\perp(\bfk)\tilde\delta_\perp^*(\bfk)\right>$, where $\tilde\delta_\perp(\bfk) = |k\hat\bfe \times \tilde\bfu|$.  \textit{Bottom right}: Cross-correlation of the phases of the density fields, defined as $\operatorname{Re}\left(\tilde\delta^*_{\rm ref}(\bfk)\tilde\delta(\bfk)\right)/\left|\tilde\delta_{\rm ref}(\bfk)\right|\left|\tilde\delta(\bfk)\right|$.
}
\label{fig:2LPT_accuracy}
\end{figure*}

The purpose of 2LPT is to correct for evolution that is missed by starting at a low redshift instead of a very high redshift.  Thus, to test the accuracy of our 2LPT theory and implementation, we run a simulation in a high-precision mode of \textsc{Abacus} (see Table \ref{tbl:abacus_params}) with $512^3$ particles in a $50\,h^{-1}$Mpc box from $z_{\rm init} = 4999$ to $z_{\rm final} = 24$.  We set up the initial conditions using the ZA in the PLT growing mode, using rescaling as described above\footnote{As a reminder, this means that, on a mode-by-mode basis, we increase/decrease the initial amplitudes to counteract the predicted linear under-/overgrowth at the target redshift.} with $z_{\rm target} = 24$.  We truncate the power spectrum at $k_{\rm Nyquist}/8$ to reduce the amplitude of the rescaling and grid effects in general to avoid non-linearities beyond those that we are seeking to measure here.  Thus, our three test cases are produced on $64^3$ particle grids, so they sample the same modes as the $512^3$ reference.  Fig.~\ref{fig:2LPT_accuracy} shows the results of comparing the following fields:
\begin{enumerate}[label=(\roman*)]
\item the Zel'dovich Approximation (labeled ``No 2LPT'');
\item \textsc{2LPTic} \citep{Crocce+2006}, a standard Fourier-space 2LPT code (``\textsc{2LPTic}'');
\item our configuration-space 2LPT (``\textsc{Abacus} 2LPT''); and
\item the reference simulation evolved from $z=4999$, as described above (``Full evolution'').
\end{enumerate}
Both \textsc{Abacus} 2LPT and \textsc{2LPTic} do a very good job of reproducing the full evolution -- much better than ZA in all metrics.  Indeed, on large scales (larger than $\sim\kNy/3$), \textsc{Abacus} 2LPT and \textsc{2LPTic} are nearly indistinguishable, and only show systematic differences of $<0.5\%$ from the reference solution.

On smaller scales, the most discriminatory metrics are the root-mean-square scatter of $P(\bfk)$ (Fig.~\ref{fig:2LPT_accuracy}, top right) and the transverse power $P_{\perp\hat\bfe}(k)$ (bottom left).  The former is useful since $k$-averaged power $P(k)$ can hide anisotropic (e.g.~lattice) effects, while the latter is useful since $P(k)$ is blind to the presence of curl modes.  At $\kNy$, \textsc{Abacus} 2LPT has 4\% scatter about the reference, while \textsc{2LPTic} has 1.5\%, since \textsc{2LPTic} does not suffer from the same anisotropic particle discreteness effects as \textsc{Abacus} (see \S\ref{sec:2lpt_theory} for a discussion).  At $\kNy/2$, these effects reduce to 1\% and 0.5\%, respectively.

In transverse power $P_{\perp\hat\bfe}(k)$, \textsc{2LPTic} reaches 8\% of its power in transverse modes at $\kNy$, versus 0\% for \textsc{Abacus} 2LPT.  Specifically, this transverse power is measured relative to the PLT eigenmodes, which are the proper eigenmodes that will not excite lattice transients.  Assuming these transverse modes do not source forces, their power will decay as $1/a$.  However, this is not a good assumption -- the non-zero eigenvalues on the lower branches in Fig.~\ref{fig:spectrum} are direct evidence of that -- which means that most of these modes will grow or oscillate indefinitely.  This means that 92\% of the continuum power at $\kNy$ is in the correct eigenmode, where it grows at a moderately suppressed rate, and 8\% is in a transverse eigenmode, where it grows at a drastically suppressed rate.  Furthermore, in all eigenmodes, the velocities (which are derived from continuum theory) will mix growing and decaying solutions.  \textsc{Abacus}'s use of PLT eigenmodes eliminates all mixing of eigenmodes and decaying solutions.

\textsc{2LPTic} operates in Fourier space, while \textsc{Abacus} operates in configuration space, so comparing the two is useful and important test.  Our detailed approach to generating comparable fields between \textsc{Abacus} and \textsc{2LPTic} is the following:
\begin{enumerate}
\item Generate a ZA field and its corresponding 2LPT field in \textsc{2LPTic};
\item Project a copy of the ZA field onto the PLT grid modes;
\item Generate a 2LPT prediction with \textsc{Abacus} from the projected ZA field;
\item Compare to the \textsc{2LPTic} 2LPT field.
\end{enumerate}
We run \textsc{2LPTic} with $\texttt{Nmesh}=512$ and $\texttt{Nsample}=64$ to generate both the $512^3$ and $64^3$ lattices.

Finally, we compare the 2LPT results from \textsc{Abacus}'s high-precision configuration to those from \textsc{Abacus}'s normal-precision configuration, since that is how it will be used in practice.  The results match to remarkable precision: the particle-averaged fractional error is $7\e{-6}$, or close to floating-point precision, which is the floor, since \textsc{Abacus}'s normal-accuracy calculations are in single precision.

\subsection{Implementation caveats}\label{sec:2lpt_caveats}
For several reasons, this ``displacement flipping'' technique would not be well-suited to a normal $N$-body code with standard Zel'dovich approximation ICs.  First, the accuracy of the 2LPT correction is a direct function of the force accuracy of the code being used.  The highly symmetric configuration of the particles makes this a particularly difficult task, because the near- and far-field components of the force both have large amplitudes (but opposite signs).  If the code does not respect the symmetry of the system, it is unlikely to produce accurate 2LPT corrections.  \textsc{Abacus} has exceptional force accuracy, even in this difficult configuration: in \textsc{Abacus}'s high-precision mode (see Table \ref{tbl:abacus_params}), a homogeneous lattice has a maximum absolute force error of $1\e{-10}$ (mean $5\e{-12}$), compared to the mean amplitude of $2\e{-3}$ for typical second-order forces at $z=49$.  In \textsc{Abacus}'s normal-precision configuration, the maximum noise is $2\e{-5}$ (mean $1\e{-6}$).  Thus, we might place an extremely conservative estimate of 1\% 2LPT errors due to noise in the lattice; in practice, however, as we showed above, the normal precision 2LPT result matches the high-precision result to an average fractional error of $7\e{-6}$.  Thus, we conclude that with \textsc{Abacus}, even our normal precision results are more than adequate to produce 2LPT corrections.\footnote{Note that we quote force errors here on a homogeneous lattice, as opposed to the random particle configuration quoted in Table \ref{tbl:abacus_params}.}

The second implementation challenge for most ICs is that the displacements should be in the longitudinal eigenmodes of the grid (see \S\ref{sec:spatial}).  Otherwise, the 2LPT corrections themselves will contain transverse modes.  Since 2LPT corrects for missing evolution from high $z$, one should not expect a system in a transient configuration that cannot be reached from high $z$ to be improved (at least on small scales) by this approach to 2LPT.  The code developed by the authors to generate initial conditions from the Zel'dovich Approximation in the PLT growing mode is publicly available\footnote{\url{https://github.com/lgarrison/zeldovich-PLT}}.

Finally, we note that our recommended implementation overwrites the particle velocities.  In theory, this is not a problem, because we can recompute the ZA velocity directly from the ZA displacement, as we prescribe in the last step of our implementation.  However, in PLT, the velocity for the pure growing solution is not related to the displacement so simply (see \S\ref{sec:temporal}).  Thus, in practice, we must either save or re-read the original first-order velocities to restore them after our 2LPT computations.

\section{Cosmological Results}\label{sec:cosmo}
We now test the impact of our modifications to the initial conditions (PLT, rescaling, and 2LPT) on common observables extracted from simulations.  Specifically, we examine the density power spectrum, halo mass function, and halo two-point correlation function.

\subsection{Simulation details}
The base cosmology for our simulations is the Planck 2015 cosmology \citep{Planck_Cosmo_2015}.  All of our simulations are initialized at $z=49$ and run to $z=1$, with outputs at $z$ = 5, 3, and 1.  Our nominal simulation size is $720^3$ particles in a $562.5\,h^{-1}$Mpc box, except for our ``oversampled'' simulation with $1440^3$ particles in the same volume.  The nominal particle mass is $4\e{10}\,h^{-1} {\rm M_\odot}$, and we use a Plummer softening length of $1/8$ of the interparticle spacing, or 78 $h^{-1}$ kpc.  Our internal code parameters are those in the ``normal precision'' column of Table \ref{tbl:abacus_params}, with $\texttt{CPD} = 225~(375)$ for $720^3~(1440^3)$ particles.

Each simulation was repeated 4 times with different realizations (``phases'') of the input power spectrum.  The results that follow are the average of the four, with error bars representing the full range of variation across the phases (i.e.~not the standard deviation).

The $1440^3$ simulation serves as our point of reference in the results below.  We truncate the input power spectrum at $\kNy/2$, so we are oversampling the existing modes in the $720^3$ boxes by a factor of two.  Thus, we consider it a more accurate representation of the ``fluid truth'' value, although it does not represent an absolutely converged reference point.  We hold the softening fixed; i.e., the softening is 1/4 of the interparticle spacing in the $1440^3$ simulation.

\subsection{Power spectrum}\label{sec:power}
We measure the density power spectrum at $z=1$ both in projection and as a full 3D set of modes.  In both cases, the density field is calculated with triangle-shaped cloud (TSC) mass assignment and deconvolved with the aliased-TSC window function from \citet[Chapter 7]{Jeong_2010}.

\subsubsection{3D power spectrum}
\begin{figure}
\centering
\includegraphics[width=\linewidth]{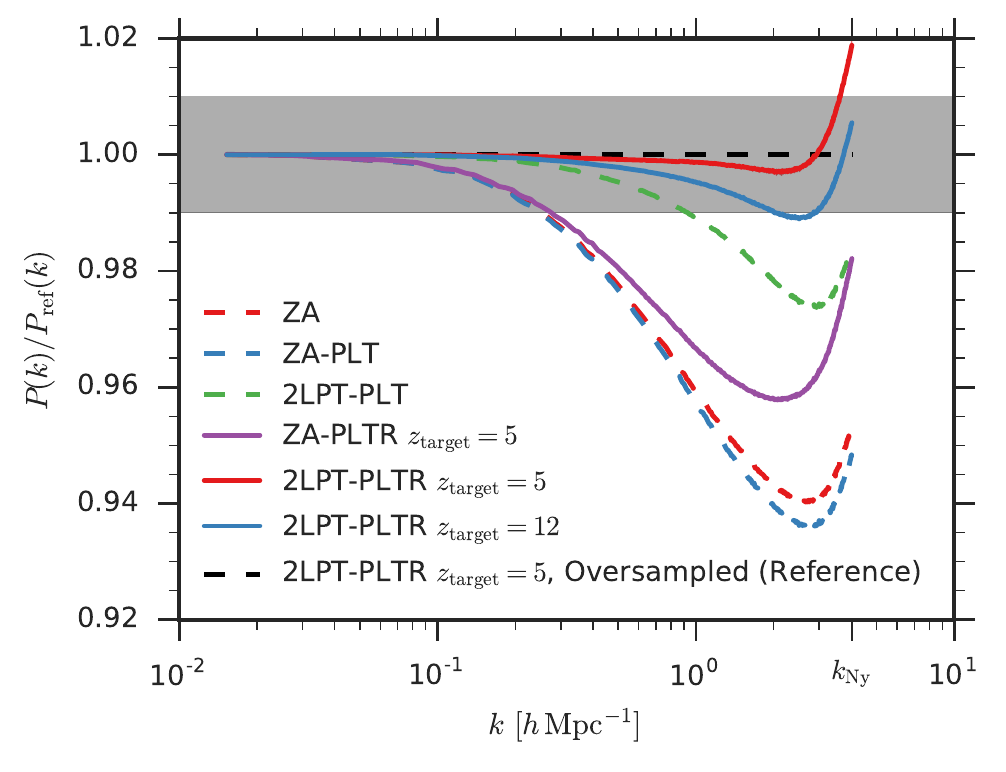}
\caption{Density power spectra at $z=1$ for different initial conditions.  Each line is the average of 4 different simulations corresponding to different realizations of the input power spectrum.  The red shaded region encloses total variability between the realizations for our preferred ICs (red solid line), but the variability is smaller than the line width.  The grey shaded band indicates our target of 1\% accuracy in the power spectrum.  See \S\ref{sec:power}.}
\label{fig:power_spectra}
\end{figure}

We compute the 3D density power spectrum with a fast Fourier transform (FFT) on a $720^3$ mesh (Fig.~\ref{fig:power_spectra}).  The combination of 2LPT and rescaling (2LPT-PLTR)  (with either $z_{\rm target} = 5$ or $12$) reproduces the power spectrum of the oversampled simulation to within 1\% for nearly the whole range of modes down to $\kNy$.  With just 2LPT, the accuracy falls below 1\% at $\kNy/4$.  In other words, to achieve 1\% accuracy at $\kNy$, a simulation with only 2LPT would have to have 64 times the mass resolution as a simulation that also uses rescaling.

All of the results in Fig.~\ref{fig:power_spectra} agree with our expectations.  The Zel'dovich Approximation (ZA) misses substantial power (between 1\% and 6\%) at all but the largest scales, and adding our PLT eigenmode corrections (ZA-PLT) slightly worsens the $z=1$ power.  This is because the ZA-PLT initial velocities are smaller to match correctly the generically suppressed growth rate.  Adding second-order Lagrangian Perturbation Theory (2LPT-PLT) is extremely helpful in recovering power on all scales, but 1 to 3\% errors persist above $\kNy/4$, corresponding roughly to 64 particle haloes.  Combining rescaling and ZA (ZA-PLTR) helps recover power on small scales, but not as well as 2LPT, and does very little on large scales.  The combination of rescaling and 2LPT (2LPT-PLTR) produces the best match to the oversampled simulation, with sub-1\% errors nearly down to $\kNy$, largely independent of our choice of $z_{\rm target}$.

Since a consideration of this work is the anisotropy of the simulation imposed by the axes of the particle lattice, we also produce a 3D power spectrum by rotating the simulation domain such that the particle lattice is skew to the FFT mesh and then measuring the 3D power spectrum.  We also shrink the domain of the FFT by a factor of $\sqrt{3}$ to avoid gaps at the edges due to the rotation.  The resulting power spectrum shows no substantial differences from Fig.~\ref{fig:power_spectra}; thus, we do not show it here.

\subsubsection{Projected power spectrum}
\begin{figure}
\centering
\includegraphics[width=\linewidth]{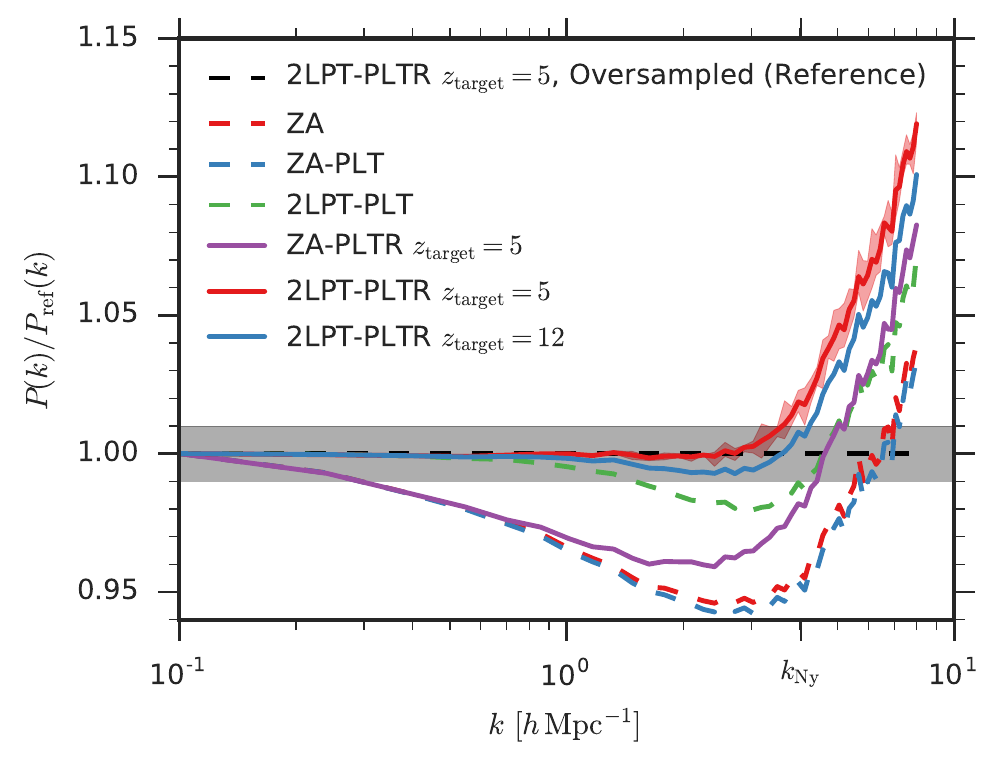}
\caption{Projected power spectra at $z=1$ for different initial conditions.  The simulation domain is rotated to an angle skew to the FFT grid before projection.  Each line is the average of 4 different simulations, each corresponding to a different realization of the input power spectrum.  The red shaded region encloses total variability between the realizations for our preferred ICs (red solid line).  The grey shaded band indicates our target of 1\% accuracy in the power spectrum.  See \S\ref{sec:power}.}
\label{fig:power_spectra_2D}
\end{figure}

Weak lensing measures the projection of the matter power spectrum on the sky, so forecasts of weak lensing from simulations must use the 2D (projected) power spectrum, which we compute with a $(8 \times 720)^2$ FFT (Fig.~\ref{fig:power_spectra_2D}).  Before projection, we rotate the simulation domain so the particle lattice is skew to the FFT mesh (as described above).  However, this makes almost no difference in the resulting power spectrum.  Note that we have plotted power above $\kNy$, where the particle lattice contributes significant power.  This should serve as a cautionary example against considering evolved power in simulations above $\kNy$; however, we do expect that the non-linearities of structure formation will somewhat lessen the amplitude of these effects at lower $z$.

\subsection{Halo mass function}
We measure halo properties with three halo finders: \textsc{rockstar} \citep{Behroozi+2013}, \textsc{rockstar} spherical overdensity (SO), and friends-of-friends (FoF).  On large scales, the three are in very good agreement concerning the behavior of the nominal-resolution simulations compared to the oversampled reference simulation, but below $\sim 500$ particles we find qualitative differences.  Understanding the behavior of different halo finders in this regime is relevant to interpreting our results -- especially the impact of rescaling, which is fundamentally a small-scale correction.

The improper growth of modes near $\kNy$ has been identified as the dominant source of systematic errors in high-precision halo mass functions \citep{Warren_2013}.  Previous attempts to correct the anisotropic aspect of this improper growth have resulted in dramatic suppression of small haloes \citep{Reed+2013} and, to our knowledge, no attempts have been made to compensate for the improper growth rates before this work.

We search for haloes at $z=1$, $3$, and $5$, and restrict our analysis to haloes larger than 30 particles and halo mass bins with approximately 100 haloes or more.

\subsubsection{\textsc{rockstar}}
\textsc{rockstar} is a hierarchical halo finder that recursively applies the friends-of-friends algorithm (see below) in six-dimensional phase space to identify dark matter structure and substructure.  \textsc{rockstar} can also track haloes across time, but we restrict our analysis to strictly static simulation snapshots.  We also only examine parent haloes (whose masses include all substructure).  A parent halo is a halo whose center does not lie within the the radius of a larger halo.  We use the default ``virial density'' threshold to define haloes.

\begin{figure*}[p]
\centering
\includegraphics[width=\textwidth]{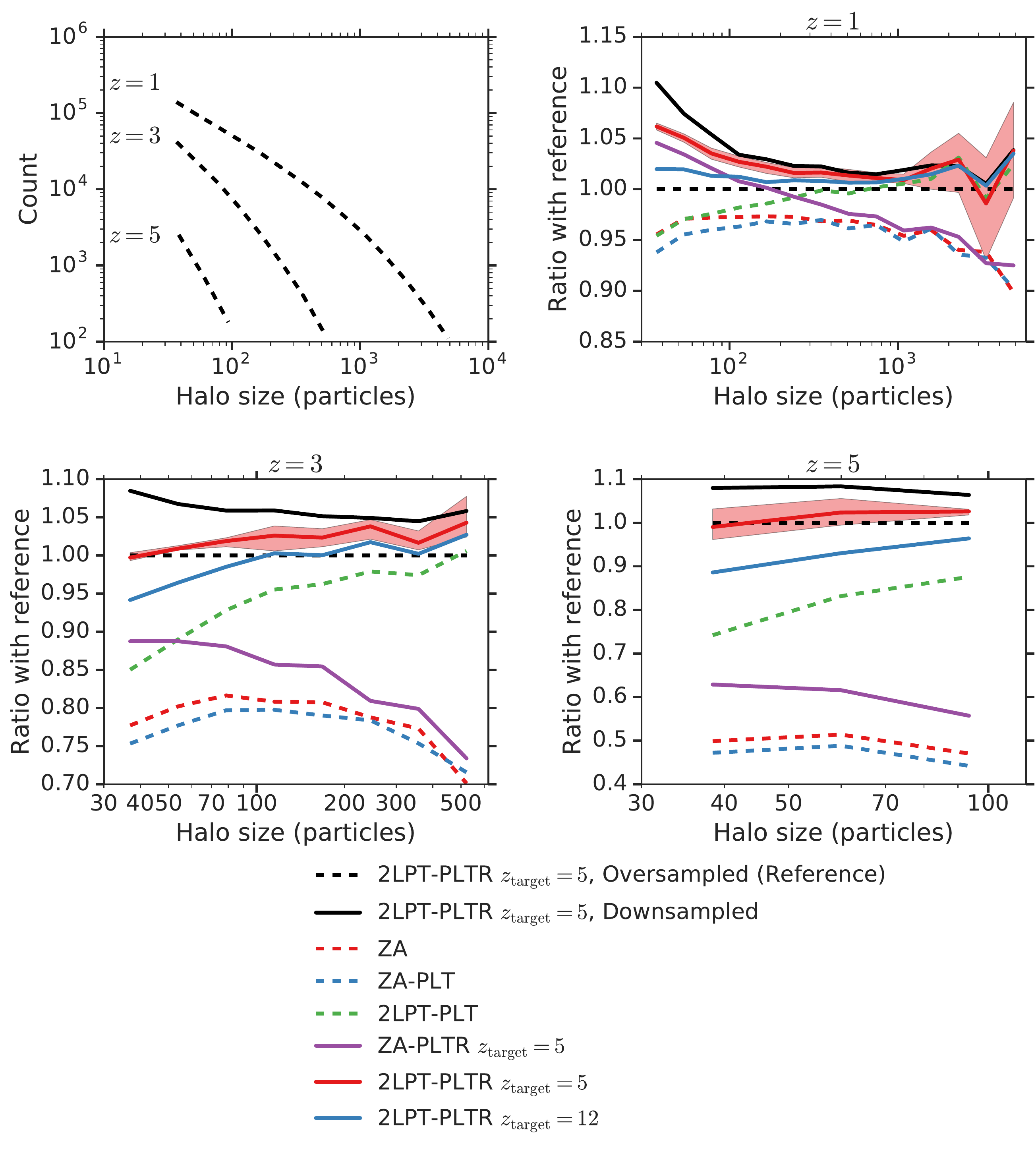}
\caption{\textit{Top left}: \textsc{rockstar} halo mass functions at the three output redshifts in the reference (oversampled) simulation.  Each line corresponds to one of the three other panels.  Halo particle counts have been divided by 8, to show them on the same mass scale as our other simulations. \textit{Top right, bottom left, bottom right}: The halo mass functions at the three output redshifts divided by the reference mass function at that redshift.  These results are summarized in Table \ref{tbl:halo_mass_rockstar}.}
\label{fig:halo_mass_rockstar}
\end{figure*}

The \textsc{rockstar} halo mass functions are shown in Fig.~\ref{fig:halo_mass_rockstar} and summarized in Table \ref{tbl:halo_mass_rockstar}.  The same trends that were visible in the $z=1$ power spectrum (\S\ref{sec:power}) are present in the halo mass function.  These effects are perhaps most clearly elucidated at $z=3$, where we find a 20 to 25\% deficit of haloes across the whole mass range of 30 -- 500 particles when using the Zel'dovich Approximation (ZA), compared to the oversampled reference simulation (corresponding to the mass range of 240 -- 4000 particles).  Adding rescaling to the initial conditions restores half of the missing small haloes, but does very little for the large haloes, as we would expect.  2LPT is the most important factor for recovering large haloes, and also has an appreciable impact on small haloes.  The combination of 2LPT and rescaling successfully recovers the halo mass function of the oversampled simulation to within 5\% across the whole mass range at this redshift.

At $z=1$, our preferred combination of 2LPT and rescaling successfully restores the 5 to 10\% deficit of haloes seen with ZA across the whole mass range of 30 -- 5000 particles.  However, below about 500 particles, we also overproduce haloes by 1 to 6\%.  To test the origin of this surplus, we downsample the $1440^3$ reference simulation by a factor of 8 and then run \textsc{rockstar} on the resulting $720^3$ particles.  Specifically, we downsample by a factor of 2 on every dimension of the original particle lattice, such that we select the particles whose initial lattice sites matches those in the $720^3$ simulations.  The result of this procedure is labeled ``Downsampled'' in Fig.~\ref{fig:halo_mass_rockstar}, where we see that downsampling tends to overproduce small haloes by 3 to 10\%.  Thus, it appears that the over-production of small haloes in our preferred ICs (rescaling and 2LPT) may be an artefact of halo finding in a coarsely sampled simulation, rather than a physical feature of the simulation itself.

Because our downsampling procedure is likely to produce non-physical variations in the binding energy of haloes, we disable halo unbinding in all of our \textsc{rockstar} analyses.  This corresponds to setting \texttt{UNBOUND\_THRESHOLD} to 0 and disabling \texttt{BOUND\_PROPS}.

\subsubsection{\textsc{rockstar} SO}
\textsc{rockstar} also has the option to output spherical overdensity masses.  The correspondence between \textsc{rockstar} haloes and SO haloes is one-to-one -- SO simply uses the halo centers that \textsc{rockstar} has already identified.  The SO mass is computed by expanding a spherical search volume from the halo center until the average density within the sphere falls below the threshold density.  SO masses are interesting for our analysis, because, naively, we would expect this technique to be relatively less sensitive to the difference in the spatial sampling between the nominal-resolution and oversampled simulations.  In particular, FoF-based techniques tend to link together haloes along filamentary structures, which SO will not do \citep{Knebe+2013}.

\begin{figure*}[p]
\centering
\includegraphics[width=\textwidth]{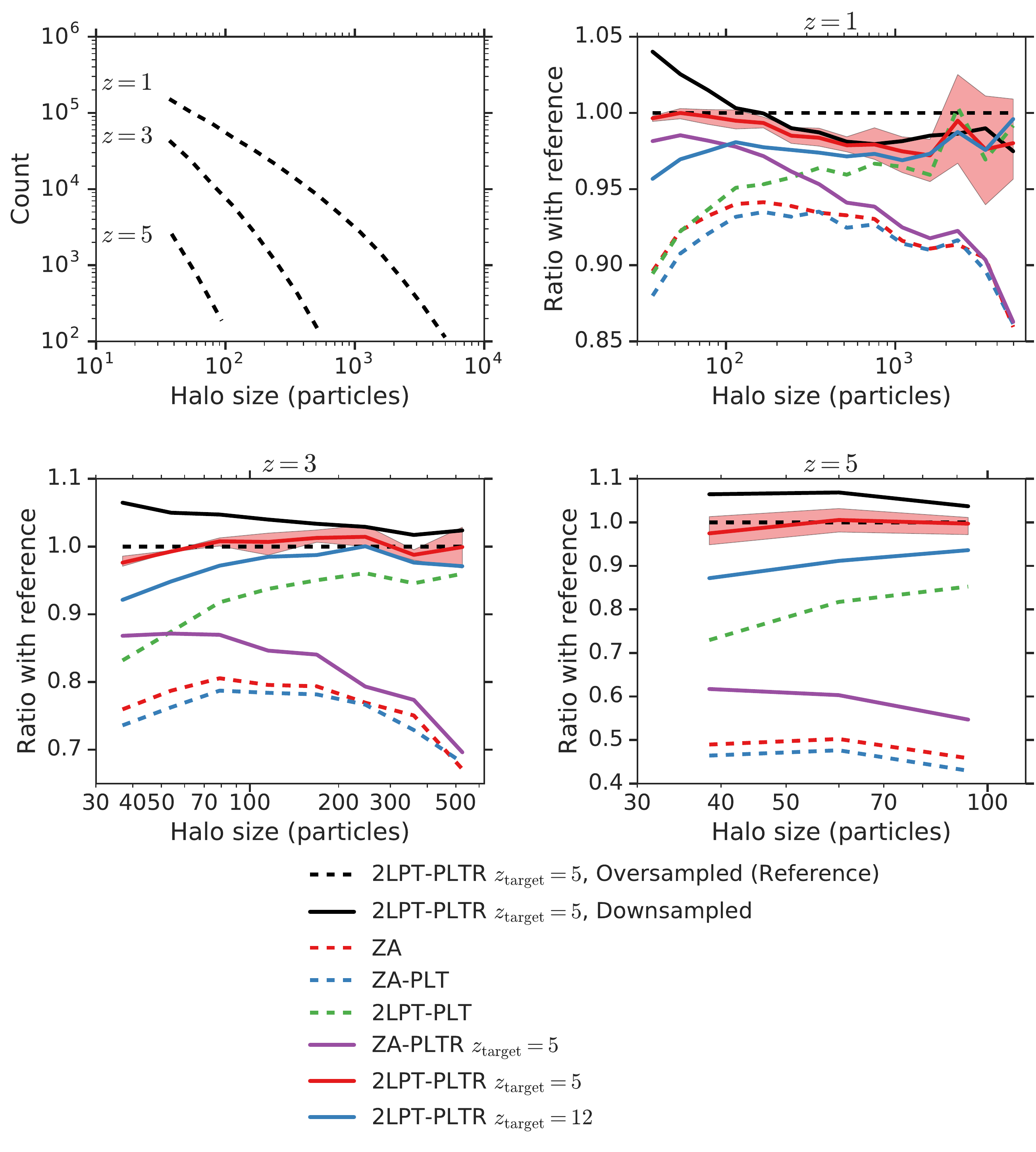}
\caption{Same as Fig.~\ref{fig:halo_mass_rockstar}, but for \textsc{rockstar} spherical overdensity halo masses.  These results are summarized in Table \ref{tbl:halo_mass_rockstar_SO}.}
\label{fig:halo_mass_rockstar_SO}
\end{figure*}

The \textsc{rockstar} SO mass functions are shown in Fig.~\ref{fig:halo_mass_rockstar_SO} and summarized in Table \ref{tbl:halo_mass_rockstar_SO}.  The same trends are visible in SO masses that were visible in \textsc{rockstar} masses.  However, while we do not overproduce small haloes at $z=1$, we do slightly underproduce large haloes at $z=1$.  This is evidence of a systematic mass shift across the whole mass range.  Specifically, on a halo-by-halo basis, switching from \textsc{rockstar} to \textsc{rockstar} SO inflates the masses of the haloes in the oversampled simulation more than the haloes in the nominal-resolution simulations.

\subsubsection{Friends-of-friends}
The friends-of-friends (FoF) algorithm links particles together if they are within a ``linking length'' $b$.  A set of linked particles is identified as a halo.  The linking length is commonly expressed as a fraction the interparticle spacing $\Delta x = N^{1/3}$.  We use $b = 0.2$ in this analysis.

\begin{figure*}[p]
\centering
\includegraphics[width=\textwidth]{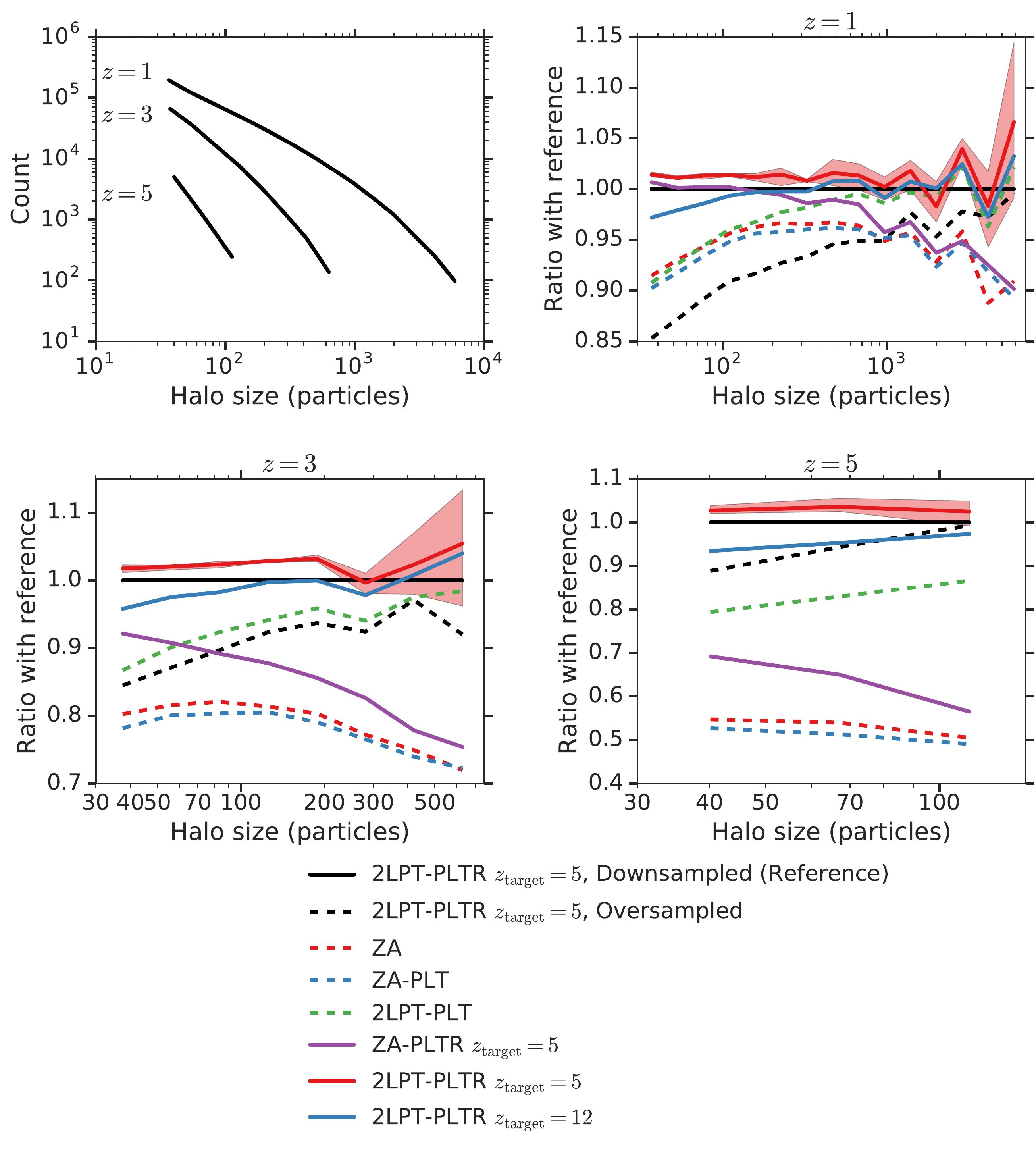}
\caption{Same as Fig.~\ref{fig:halo_mass_rockstar}, but for friends-of-friends halo masses.  Note that the reference line is the ``downsampled'' result, because the ``oversampled'' result is a very poor match to the nominal-resolution results.  The downsampled result is produced by taking a subsample of one out of eight particles and running that subsample through FoF.  This matches the spatial stochasticity of the nominal-resolution simulations and produces results that agree with the \textsc{rockstar} and power spectra results.  These results are summarized in Table \ref{tbl:halo_mass_FoF}.}
\label{fig:halo_mass_FoF}
\end{figure*}

The FoF mass functions are shown in Fig.~\ref{fig:halo_mass_FoF} and summarized in Table \ref{tbl:halo_mass_FoF}.  FoF is very sensitive to the spatial stochasticity of the particle sampling, due to the Poisson nature of particle linking and the above-mentioned filamentary linking problem.  Thus, we consider the downsampled simulation (where we take 1/8 of the oversampled particles) as our reference case.  Without this, we would conclude that our simulations overproduce haloes below 1000 particles by 5 to 15\% at $z=1$.  We consider the above \textsc{rockstar} results as further evidence that the downsampled simulation is indeed the appropriate reference case.

Our preferred ICs using rescaling and 2LPT match the downsampled mass function to within 1\% across nearly the whole mass range of 30 -- 2000 particles at $z=1$, with similar success at the other redshifts.  At high masses, the small number of haloes causes fluctuations at the 5\% level, but there is no evidence for additional systemic shifts.

The particularly simple nature of FoF makes it unsurprising that we recover very similar results to the power spectrum analysis, namely, that our preferred ICs are an excellent match to the oversampled simulation, and that 2LPT alone is insufficient below 500 particles, with 5\% too few haloes of 100 particles.

\begin{table*}
\begin{tightcenter}
\caption{Mean errors in the \textsc{rockstar} halo mass functions (Fig.~\ref{fig:halo_mass_rockstar})}
 \label{tbl:halo_mass_rockstar}
\begin{tabular}{|l|rr|rr|rr|}
\hline
 Simulation                                              & \multicolumn{2}{|c|}{$z=1$}     & \multicolumn{2}{|c|}{$z=3$}      & \multicolumn{2}{|c|}{$z=5$}      \\
\hline
                                                         & Mean                          & RMS & Mean                          & RMS  & Mean                          & RMS  \\
 2LPT-PLTR $z_{\rm target} = 5$, Oversampled (Reference) & 0.0\%                           & 0.0\% & 0.0\%                           & 0.0\%  & 0.0\%                           & 0.0\%  \\
 2LPT-PLTR $z_{\rm target} = 5$, Downsampled             & 3.4                           & 4.3 & 5.9                           & 6.0  & 7.6                           & 7.6  \\
 ZA                                                      & --4.2                          & 4.7 & --21.6                         & 21.9 & --50.6                         & 50.6 \\
 ZA-PLT                                                  & --4.8                          & 5.1 & --22.9                         & 23.1 & --53.3                         & 53.3 \\
 2LPT-PLT                                                & --0.6                          & 2.1 & --5.7                          & 7.4  & --18.3                         & 19.2 \\
 ZA-PLTR $z_{\rm target} = 5$                            & --1.7                          & 3.9 & --16.1                         & 16.9 & --39.9                         & 40.1 \\
 2LPT-PLTR $z_{\rm target} = 5$                          & 2.4                           & 3.0 & 2.2                           & 2.6  & 1.3                           & 2.1  \\
 2LPT-PLTR $z_{\rm target} = 12$                         & 1.3                           & 1.6 & --0.7                          & 2.7  & --7.3                          & 8.0  \\
\hline
\end{tabular}
\end{tightcenter}
{\sc Notes} -- Summary of the offset and scatter of the halo mass functions for different ICs relative to the reference ICs.  The ``mean'' deviation for a given simulation is the average fractional difference of that simulation's halo mass function from the reference.  The ``RMS'' is the root-mean-square of this fractional difference.
\end{table*}

\begin{table*}
\begin{tightcenter}
\caption{Mean errors in the \textsc{rockstar} SO halo mass functions (Fig.~\ref{fig:halo_mass_rockstar_SO})}
 \label{tbl:halo_mass_rockstar_SO}
\begin{tabular}{|l|rr|rr|rr|}
\hline
 Simulation                                              & \multicolumn{2}{|c|}{$z=1$}     & \multicolumn{2}{|c|}{$z=3$}      & \multicolumn{2}{|c|}{$z=5$}      \\
\hline
                                                         & Mean                          & RMS & Mean                          & RMS  & Mean                          & RMS  \\
 2LPT-PLTR $z_{\rm target} = 5$, Oversampled (Reference) & 0.0\%                           & 0.0\% & 0.0\%                           & 0.0\%  & 0.0\%                           & 0.0\%  \\
 2LPT-PLTR $z_{\rm target} = 5$, Downsampled             & --0.4                          & 1.9 & 3.8                           & 4.1  & 5.7                           & 5.8  \\
 ZA                                                      & --8.0                          & 8.3 & --23.3                         & 23.7 & --51.7                         & 51.7 \\
 ZA-PLT                                                  & --8.6                          & 8.9 & --24.7                         & 24.9 & --54.3                         & 54.4 \\
 2LPT-PLT                                                & --4.3                          & 5.0 & --7.8                          & 8.9  & --20.0                         & 20.7 \\
 ZA-PLTR $z_{\rm target} = 5$                            & --5.5                          & 6.5 & --18.0                         & 18.9 & --41.1                         & 41.2 \\
 2LPT-PLTR $z_{\rm target} = 5$                          & --1.4                          & 1.7 & --0.0                          & 1.2  & --0.8                          & 1.5  \\
 2LPT-PLTR $z_{\rm target} = 12$                         & --2.5                          & 2.6 & --3.0                          & 3.8  & --9.4                          & 9.7  \\
\hline
\end{tabular}
\end{tightcenter}
{\sc Notes} -- See Table \ref{tbl:halo_mass_rockstar} Notes.
\end{table*}

\begin{table*}
\begin{tightcenter}
\caption{Mean errors in the friends-of-friends halo mass functions (Fig.~\ref{fig:halo_mass_FoF})}
 \label{tbl:halo_mass_FoF}
\begin{tabular}{|l|rr|rr|rr|}
\hline
 Simulation                                              & \multicolumn{2}{|c|}{$z=1$}     & \multicolumn{2}{|c|}{$z=3$}      & \multicolumn{2}{|c|}{$z=5$}      \\
\hline
                                                         & Mean                          & RMS & Mean                          & RMS  & Mean                          & RMS  \\
 2LPT-PLTR $z_{\rm target} = 5$, Downsampled (Reference) & 0.0\%                           & 0.0\% & 0.0\%                           & 0.0\%  & 0.0\%                           & 0.0\%  \\
 2LPT-PLTR $z_{\rm target} = 5$, Oversampled             & --6.5                          & 7.6 & --8.9                          & 9.6  & --5.8                          & 7.2  \\
 ZA                                                      & --5.6                          & 6.1 & --21.3                         & 21.5 & --46.9                         & 47.0 \\
 ZA-PLT                                                  & --6.1                          & 6.5 & --22.4                         & 22.6 & --49.0                         & 49.0 \\
 2LPT-PLT                                                & --2.5                          & 3.9 & --6.3                          & 7.3  & --17.0                         & 17.3 \\
 ZA-PLTR $z_{\rm target} = 5$                            & --2.7                          & 4.1 & --14.8                         & 15.9 & --36.4                         & 36.8 \\
 2LPT-PLTR $z_{\rm target} = 5$                          & 1.4                           & 2.4 & 2.5                           & 2.9  & 2.9                           & 3.0  \\
 2LPT-PLTR $z_{\rm target} = 12$                         & --0.2                          & 1.7 & --0.8                          & 2.5  & --4.7                          & 4.9  \\
\hline
\end{tabular}
\end{tightcenter}
{\sc Notes} -- See Table \ref{tbl:halo_mass_rockstar} Notes.
\end{table*}

\subsection{Halo clustering}
We compute the two-point correlation function (2PCF) of haloes as
\begin{align}
\xi(s) = \frac{DD}{RR} - 1,
\end{align}
which is equivalent to the Landy-Szalay estimator \citep{Landy_Szalay_1993} because the domain is a cubic, periodic box.  We restrict our analysis to haloes separated by 3 -- 30 $h^{-1}$Mpc at $z=1$, where we have a sufficient number of halo pairs and are not approaching the box periodicity scale.  We compute the $DD$ term using {\sc TreeCorr} \citep{TreeCorr} and the $RR$ term analytically.  We repeat this analysis for the haloes found by each halo finder.

We compute the correlation function in four mass bins.  The first two are simple halo mass cuts: 100 -- 300 particles, and 300 -- 1000 particles.  The next two are halo-mass rank cuts: all haloes ranked between $10^5$ and $3 \times 10^4$, and all haloes ranked above $3 \times 10^4$.  These ranks are chosen such that the average mass limits across phases and ICs approximately correspond to the absolute mass cuts of the first two bins.  This halo-mass ranking procedure is akin to ``abundance matching'', wherein dark matter haloes are matched to their observational counterparts by matching the mass rank of each, rather than attempting to make absolute mass calibrations \citep[e.g.][]{Guo+2010}.  This is particularly important for volume-limited galaxy surveys, which have no direct knowledge of the masses of haloes but do know the number densities of the most massive haloes.  The following results suggest that clustering derived from abundance matching has a relatively smaller systematic error due to details of the initial conditions than clustering from a mass-selected set of haloes.

\subsubsection{\textsc{rockstar}}
\begin{figure*}
\centering
\includegraphics[width=\textwidth]{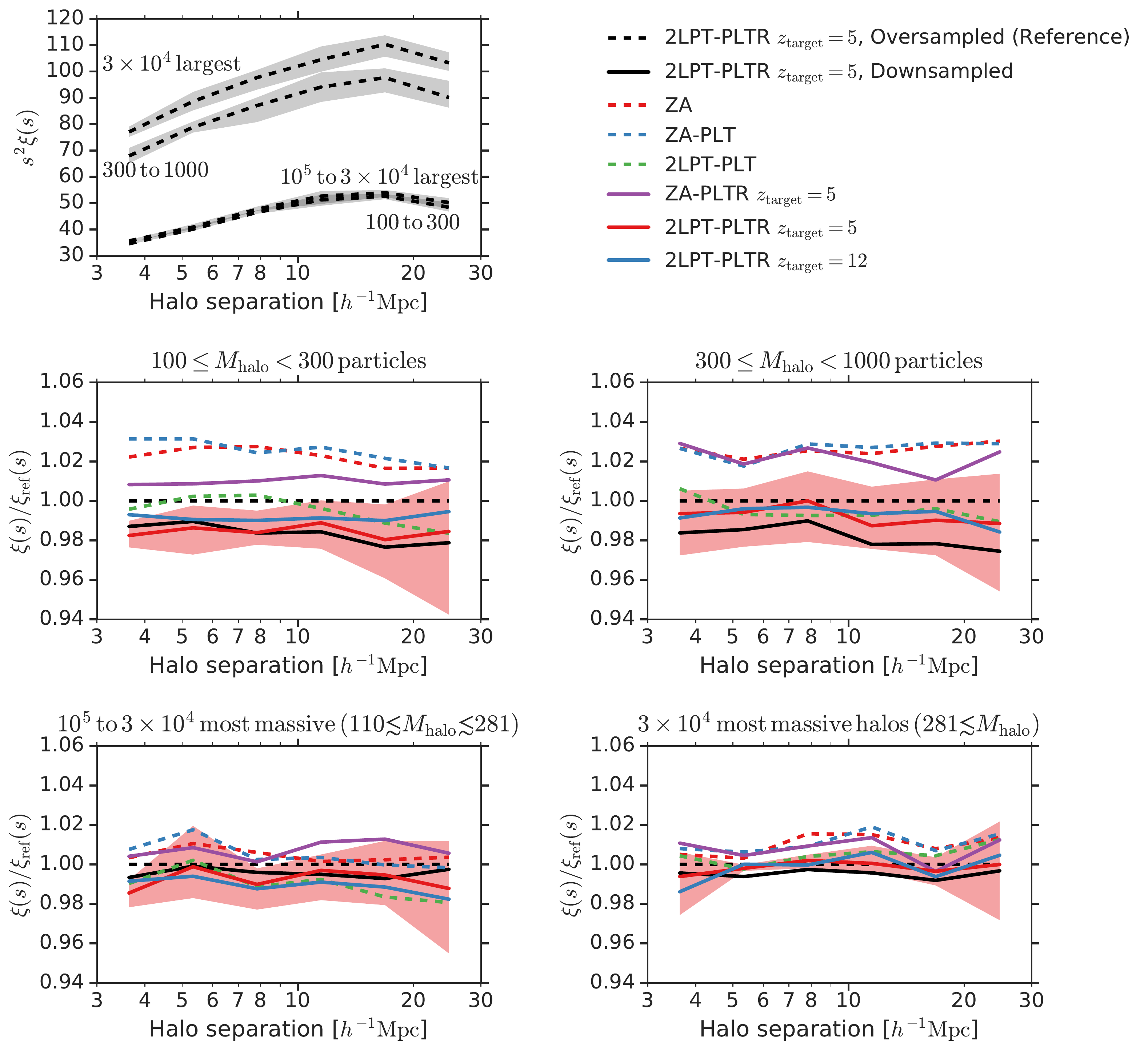}
\caption{The two-point correlation function of \textsc{rockstar} haloes at $z=1$.  Each of the bottom four panels shows a different mass bin.  The middle row shows mass-selected haloes, while the bottom row shows mass-rank-selected.  The approximate mass ranges are the average across all of the simulations.  The average error incurred by using each of these initial conditions is summarized in Table \ref{tbl:2pcf_rockstar}.}
\label{fig:2pcf_rockstar}
\end{figure*}

The 2PCF of \textsc{rockstar} haloes is shown in Fig.~\ref{fig:2pcf_rockstar} and summarized in Table \ref{tbl:2pcf_rockstar}.  The absolute-mass bins show large sensitivity ($\sim 5\%$) to the ICs, because the masses of haloes at a constant bias are systematically shifting up or down.  Thus, the mass bins are gaining or losing some bias relative to the reference simulation, shifting the 2PCF.

Switching to mass-ranked bins reduces the scatter among the ICs by a factor of 2 or more.  Regardless of absolute-mass or mass-ranked binning, either 2LPT or 2LPT with rescaling is the best match to the reference.  With mass ranking, we can recover the 2PCF to within a fraction of a percent, especially after averaging over phases.  Adding volume to our boxes would produce the same effect, which is one reason why multi-Gpc boxes will be required to calibrate upcoming galaxy surveys.

The most important factor for recovering the 2PCF is 2LPT, followed by rescaling.  Without 2LPT, systematic shifts of 1 to 3\% are seen in all absolute-mass bins, or 1\% in the mass-ranked analyses.

\subsubsection{\textsc{rockstar} SO}
\begin{figure*}
\centering
\includegraphics[width=\textwidth]{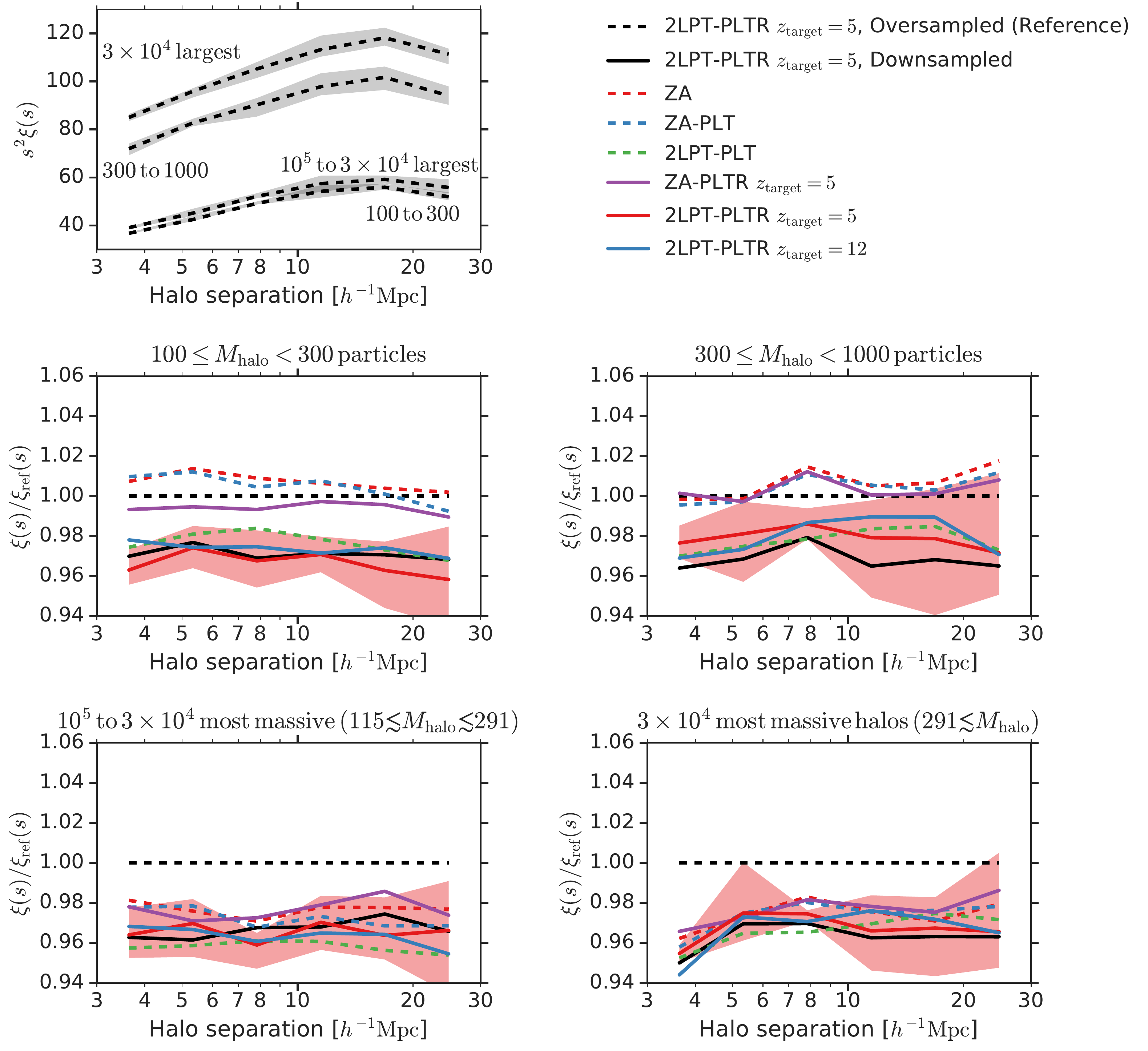}
\caption{Same as Fig.~\ref{fig:2pcf_rockstar}, except with \textsc{rockstar} spherical overdensity halo masses.  These results are summarized in Table \ref{tbl:2pcf_rockstar_SO}.}
\label{fig:2pcf_rockstar_SO}
\end{figure*}

The 2PCF of \textsc{rockstar} SO haloes is shown in Fig.~\ref{fig:2pcf_rockstar_SO} and summarized in Table \ref{tbl:2pcf_rockstar_SO}.  We see the same reduction in scatter moving from absolute-mass to mass-ranked bins, but we see a large (3\%) offset from the reference solution that was not present before.  Since the halo centers are identical to the \textsc{rockstar} halo centers, the SO haloes must be reordering the masses of haloes, such that the $3 \times 10^4$ most massive haloes (for example) are a relatively less-biased sample than the reference.  We note that the downsampled simulation is an excellent match to the nominal-resolution simulations, as it is in the mass functions, suggesting that mass-reordering due to the sensitivity of \textsc{rockstar} to the mass resolution of the simulations is the main culprit.

\subsubsection{Friends-of-friends}
\begin{figure*}
\centering
\includegraphics[width=\textwidth]{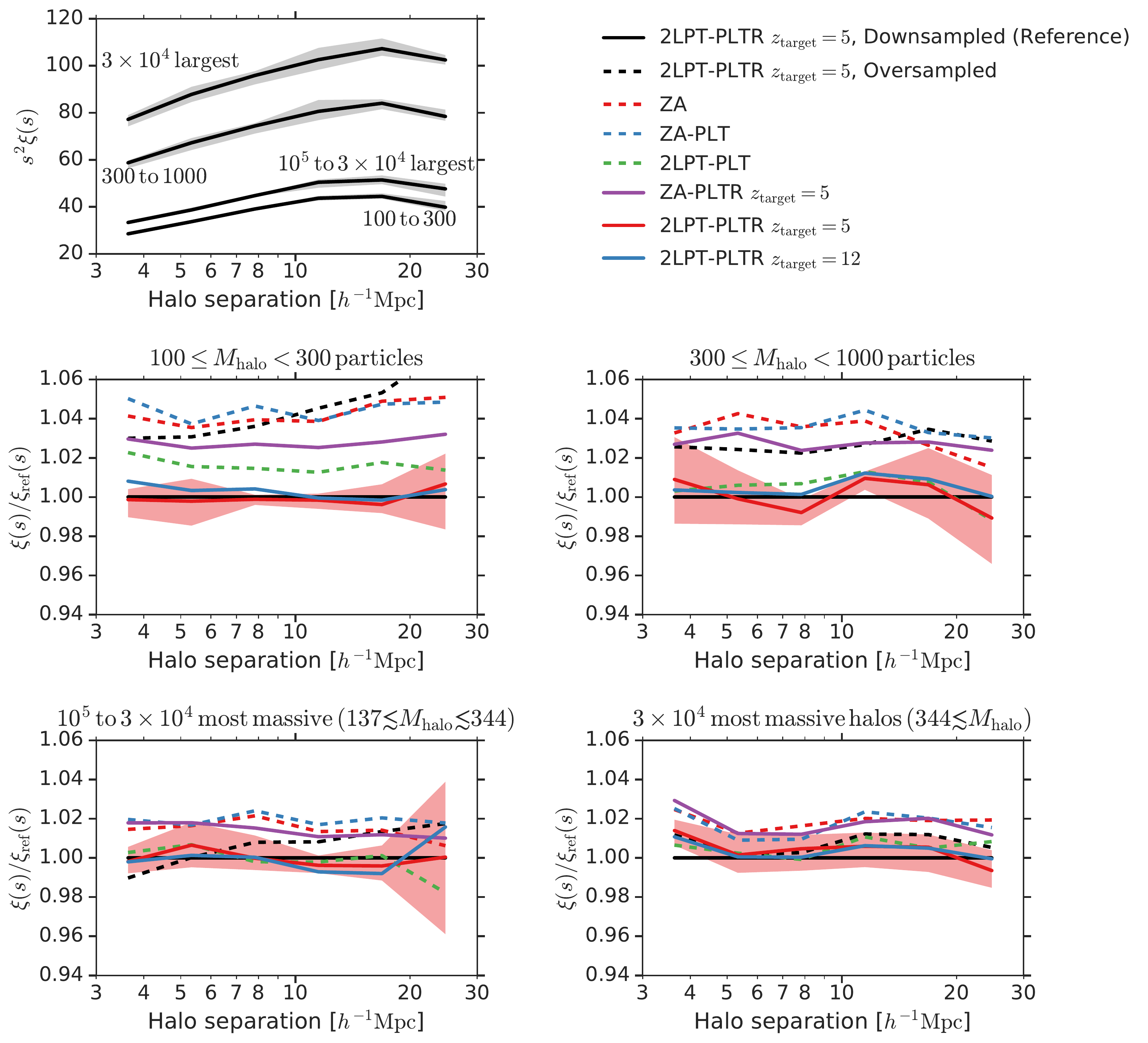}
\caption{Same as Fig.~\ref{fig:2pcf_rockstar}, except with friends-of-friends halo masses.  These results are summarized in Table \ref{tbl:2pcf_FoF}.}
\label{fig:2pcf_FoF}
\end{figure*}

The 2PCF of FoF haloes is shown in Fig.~\ref{fig:2pcf_FoF} and summarized in Table \ref{tbl:2pcf_FoF}.  The FoF results are quite similar to the \textsc{rockstar} results (with 2LPT and rescaling offering the best match to the reference, followed by 2LPT alone), with the exception that our preferred solution works equally well in the absolute-mass and mass-ranked bins.  This is due to the excellent match in the mass function between the reference and the 2LPT results, so very little halo reordering must occur when switching to the mass-ranked functions.  Our preferred solution recovers the 2PCF to within a fraction of a percent in most cases. The other ICs are shifted by 2 to 4\% in the absolute-mass bins, or 1 to 2\% in the mass-ranked.

\begin{table*}
\caption{Mean errors in the \textsc{rockstar} halo two-point correlation functions (Fig.~\ref{fig:2pcf_rockstar})}
\label{tbl:2pcf_rockstar}
\begin{tightcenter}
\begin{tabular}{|l|rr|rr|rr|rr|}
\hline
Simulation                                              & \multicolumn{2}{|c|}{\begin{tabular}{@{}c@{}} $100$ -- $300$ \\ particles \end{tabular}}     & \multicolumn{2}{|c|}{\begin{tabular}{@{}c@{}} $300$ -- $1000$ \\ particles \end{tabular}}     & \multicolumn{2}{|c|}{\begin{tabular}{@{}c@{}} $10^5$ to $3\times 10^4$ \\ most massive \end{tabular}}     & \multicolumn{2}{|c|}{\begin{tabular}{@{}c@{}} $3\times 10^4$ \\ most massive \end{tabular}}     \\
\hline

                                                         & Mean                                   & RMS & Mean                                    & RMS & Mean                                                           & RMS & Mean                                                 & RMS \\
 2LPT-PLTR $z_{\rm target} = 5$, Oversampled (Reference) & 0.0\%                                    & 0.0\% & 0.0\%                                     & 0.0\% & 0.0\%                                                            & 0.0\% & 0.0\%                                                  & 0.0\% \\
 2LPT-PLTR $z_{\rm target} = 5$, Downsampled             & --1.7                                   & 1.7 & --1.8                                    & 1.9 & --0.4                                                           & 0.5 & --0.5                                                 & 0.5 \\
 ZA                                                      & 2.2                                    & 2.3 & 2.6                                     & 2.6 & 0.5                                                            & 0.6 & 1.0                                                  & 1.1 \\
 ZA-PLT                                                  & 2.5                                    & 2.6 & 2.6                                     & 2.7 & 0.5                                                            & 0.8 & 1.1                                                  & 1.2 \\
 2LPT-PLT                                                & --0.5                                   & 0.9 & --0.5                                    & 0.7 & --1.0                                                           & 1.2 & 0.5                                                  & 0.7 \\
 ZA-PLTR $z_{\rm target} = 5$                            & 1.0                                    & 1.0 & 2.2                                     & 2.2 & 0.7                                                            & 0.8 & 0.8                                                  & 1.0 \\
 2LPT-PLTR $z_{\rm target} = 5$                          & --1.6                                   & 1.6 & --0.8                                    & 0.9 & --0.8                                                           & 0.9 & --0.1                                                 & 0.3 \\
 2LPT-PLTR $z_{\rm target} = 12$                         & --0.8                                   & 0.9 & --0.7                                    & 0.8 & --1.1                                                           & 1.1 & --0.2                                                 & 0.7 \\
\hline
\end{tabular}\end{tightcenter}
{\sc Notes} -- Summary of the offset and scatter of the correlation functions for different ICs relative to the reference ICs.  The ``mean'' deviation for a given simulation is the average fractional difference of that simulation's correlation function from the reference.  The ``RMS'' is the root-mean-square of this fractional difference.
\end{table*}

\begin{table*}
\caption{Mean errors in the \textsc{rockstar} spherical overdensity halo two-point correlation functions (Fig.~\ref{fig:2pcf_rockstar_SO})}
\label{tbl:2pcf_rockstar_SO}
\begin{tightcenter}
\begin{tabular}{|l|rr|rr|rr|rr|}
\hline
Simulation                                              & \multicolumn{2}{|c|}{\begin{tabular}{@{}c@{}} $100$ -- $300$ \\ particles \end{tabular}}     & \multicolumn{2}{|c|}{\begin{tabular}{@{}c@{}} $300$ -- $1000$ \\ particles \end{tabular}}     & \multicolumn{2}{|c|}{\begin{tabular}{@{}c@{}} $10^5$ to $3\times 10^4$ \\ most massive \end{tabular}}     & \multicolumn{2}{|c|}{\begin{tabular}{@{}c@{}} $3\times 10^4$ \\ most massive \end{tabular}}     \\
\hline
                                                         & Mean                                   & RMS & Mean                                    & RMS & Mean                                                           & RMS & Mean                                                 & RMS \\
 2LPT-PLTR $z_{\rm target} = 5$, Oversampled (Reference) & 0.0\%                                    & 0.0\% & 0.0\%                                     & 0.0\% & 0.0\%                                                            & 0.0\% & 0.0\%                                                  & 0.0\% \\
 2LPT-PLTR $z_{\rm target} = 5$, Downsampled             & --2.9                                   & 2.9 & --3.2                                    & 3.2 & --3.3                                                           & 3.4 & --3.7                                                 & 3.8 \\
 ZA                                                      & 0.7                                    & 0.8 & 0.7                                     & 1.0 & --2.3                                                           & 2.3 & --2.6                                                 & 2.7 \\
 ZA-PLT                                                  & 0.5                                    & 0.8 & 0.4                                     & 0.7 & --2.8                                                           & 2.8 & --2.6                                                 & 2.7 \\
 2LPT-PLT                                                & --2.4                                   & 2.4 & --2.3                                    & 2.3 & --4.2                                                           & 4.2 & --3.4                                                 & 3.4 \\
 ZA-PLTR $z_{\rm target} = 5$                            & --0.6                                   & 0.6 & 0.3                                     & 0.6 & --2.3                                                           & 2.4 & --2.3                                                 & 2.4 \\
 2LPT-PLTR $z_{\rm target} = 5$                          & --3.4                                   & 3.4 & --2.1                                    & 2.2 & --3.5                                                           & 3.5 & --3.3                                                 & 3.3 \\
 2LPT-PLTR $z_{\rm target} = 12$                         & --2.6                                   & 2.7 & --2.0                                    & 2.2 & --3.7                                                           & 3.7 & --3.3                                                 & 3.5 \\
\hline
\end{tabular}
\end{tightcenter}
{\sc Notes} -- See Table \ref{tbl:2pcf_rockstar} Notes.
\end{table*}

\begin{table*}
\caption{Mean errors in the friends-of-friends halo two-point correlation functions (Fig.~\ref{fig:2pcf_FoF})}
\label{tbl:2pcf_FoF}
\begin{tightcenter}
\begin{tabular}{|l|rr|rr|rr|rr|}
\hline
Simulation                                              & \multicolumn{2}{|c|}{\begin{tabular}{@{}c@{}} $100$ -- $300$ \\[-2pt] particles \end{tabular}}     & \multicolumn{2}{|c|}{\begin{tabular}{@{}c@{}} $300$ -- $1000$ \\[-2pt] particles \end{tabular}}     & \multicolumn{2}{|c|}{\begin{tabular}{@{}c@{}} $10^5$ to $3\times 10^4$ \\[-2pt] most massive \end{tabular}}     & \multicolumn{2}{|c|}{\begin{tabular}{@{}c@{}} $3\times 10^4$ \\[-2pt] most massive \end{tabular}}     \\
\hline
                                                         & Mean                                   & RMS & Mean                                    & RMS & Mean                                                           & RMS & Mean                                                 & RMS \\
 2LPT-PLTR $z_{\rm target} = 5$, Downsampled (Reference) & 0.0\%                                    & 0.0\% & 0.0\%                                     & 0.0\% & 0.0\%                                                            & 0.0\% & 0.0\%                                                  & 0.0\% \\
 2LPT-PLTR $z_{\rm target} = 5$, Oversampled             & 4.5                                    & 4.8 & 2.7                                     & 2.7 & 0.6                                                            & 1.1 & 0.8                                                  & 0.9 \\
 ZA                                                      & 4.2                                    & 4.3 & 3.2                                     & 3.3 & 1.4                                                            & 1.5 & 1.9                                                  & 1.9 \\
 ZA-PLT                                                  & 4.5                                    & 4.5 & 3.5                                     & 3.6 & 1.9                                                            & 1.9 & 1.7                                                  & 1.8 \\
 2LPT-PLT                                                & 1.6                                    & 1.7 & 0.4                                     & 0.9 & --0.2                                                           & 0.8 & 0.5                                                  & 0.7 \\
 ZA-PLTR $z_{\rm target} = 5$                            & 2.8                                    & 2.8 & 2.7                                     & 2.7 & 1.4                                                            & 1.4 & 1.7                                                  & 1.8 \\
 2LPT-PLTR $z_{\rm target} = 5$                          & --0.1                                   & 0.3 & 0.1                                     & 0.8 & --0.1                                                           & 0.4 & 0.4                                                  & 0.7 \\
 2LPT-PLTR $z_{\rm target} = 12$                         & 0.3                                    & 0.4 & 0.5                                     & 0.7 & 0.0                                                            & 0.8 & 0.4                                                  & 0.5 \\
\hline
\end{tabular}
\end{tightcenter}
{\sc Notes} -- See Table \ref{tbl:2pcf_rockstar} Notes.
\end{table*}

\subsection{Glass initial conditions}\label{sec:glass}
All of the simulations in this work have used particle-lattice pre-initial conditions, because less structured configurations (such as a ``glass'') are harder to treat in the PLT framework.  Analytically, the eigenmodes are no longer plane waves, and computationally, the $N$ $3\times3$ eigenvalue problems become a $3N\times3N$ problem.  Our approach here is to generate glass initial conditions and empirically demonstrate that they do not alleviate the systematic suppression of small-scale power that is predicted by PLT.  This reproduces the result of \cite{Joyce+2009}.

A glass is a force-free configuration of particles that is reached by evolving a random distribution of particles in an expanding background with the sign of gravitational acceleration reversed.  All the particles repel each other and oscillate about their equilibrium positions until they have been sufficiently damped by the background expansion.  Like a lattice, a glass is uniform on large scales, but unlike a lattice, it is also isotropic on scales appreciably larger than the particle spacing.

We expect that glass pre-initial conditions will eliminate large-scale anisotropy produced by the particle lattice, but not the systematic small-scale suppression of power.  This effect depends only on the fact that the continuum density field has been discretized, not the nature of that discretization.  This is what \cite{Joyce_Marcos_2007b} call \textit{dynamical sparse sampling effects}.

To test this, we use \textsc{Abacus} to generate $72^3$ and $80^3$ particle glasses, which are tiled using \textsc{2LPTic} \citep{Crocce+2006} to produce $360^3$ and $720^3$ 2LPT initial conditions, respectively.  We use \textsc{2LPTic}'s cloud-in-cell deconvolution to reduce the suppression power from interpolating to the glass.  The number of \textsc{Abacus} cells-per-glass-tile is kept constant between the glass generation and actual simulations, as are the precision and softening.  This ensures no discontinuities in the residual forces as we transition from glass-making to simulation.  The mean residual forces after glass generation are a factor of 500 below the mean forces on the particles at $z_{\rm init} = 49$.

\begin{figure}
\centering
\includegraphics[width=\linewidth]{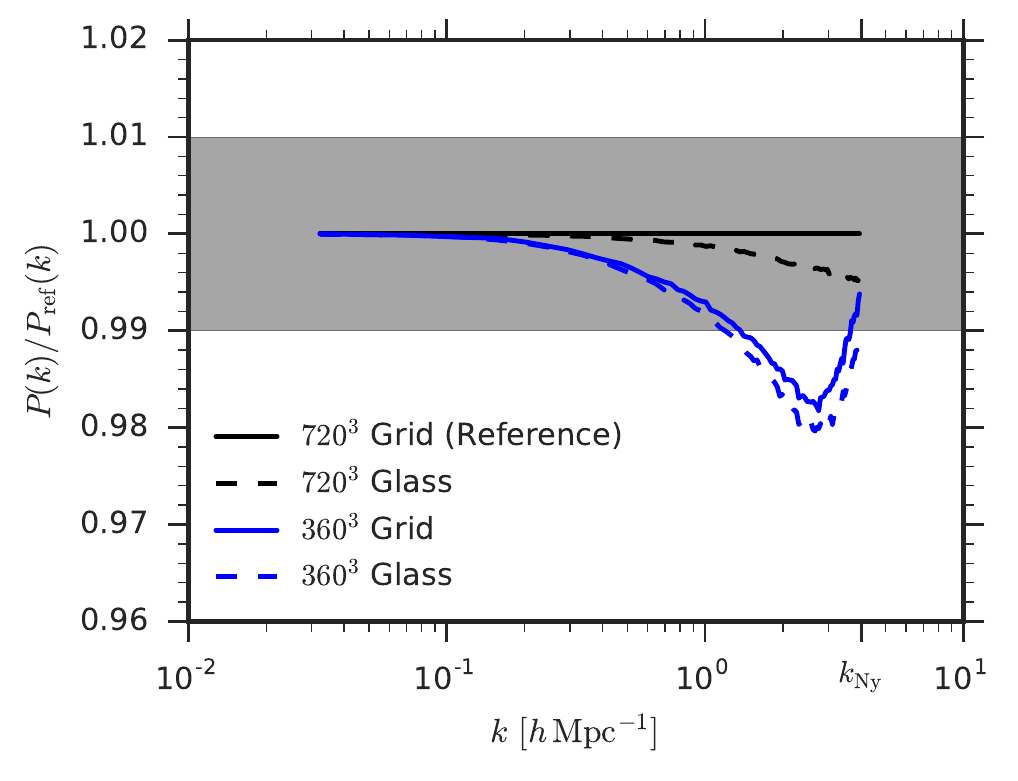}
\caption{Power spectra at $z=1$ for different pre-initial conditions.  The grey shaded band indicates our target of 1\% accuracy in the power spectrum.  The $360^3$ glass shows as much loss of power relative to the $720^3$ glass as the $360^3$ lattice does to the $720^3$ lattice.  In other words, glass initial conditions do not alleviate the small-scale suppression of power predicted by PLT.  This suppression is due to the fact that the continuum density field is discretized, not the arrangement (glass or lattice) of the discretization. See \S\ref{sec:glass}.}
\label{fig:glass}
\end{figure}

Fig.~\ref{fig:glass} shows that switching from lattice to glass pre-ICs does not restore power.  The $360^3$ glass configuration at $z=1$ shows just as much loss of power relative to the $720^3$ glass as the $360^3$ lattice does from the $720^3$ lattice.

The $720^3$ glass shows a small loss of power ($0.5\%$ at $\kNy$) relative to the $720^3$ lattice, which we attribute to loss of power during interpolation of the displacements from the FFT mesh to the particles (despite our use of CIC deconvolution\footnote{See \cite{Jenkins_2010} for a superior method of interpolating to a glass.}).  If the effect were due to residual forces in the glass tiles, we would have expected accelerated growth of structure and a surplus of power, not a deficit.  Indeed, increasing the size of the FFT mesh in \textsc{2LPTic} has no effect on our results other than to decrease this loss of power, hence our use of a relatively fine $1440^3$ mesh.

\section{Conclusions}\label{sec:conc}
We have built upon the particle linear theory of \cite{Marcos+2006} and shown how to eliminate transients at linear order in initial conditions for cosmological simulations.  We consider the case of a simple cubic lattice of particles as the pre-initial configuration instead of a glass, because these dynamical discreteness effects are present in both cases but only analytically tractable in the particle lattice case.  We then consider how such a system evolves in time and reproduce the PLT result that modes near $\kNy$ will be systematically suppressed as a simulation evolves.  PLT gives the exact amplitude of this $\bfk$-dependent suppression, so we rescale the initial mode amplitudes such that they will arrive at the correct amplitudes at a later time, with the motivation of seeding non-linear evolution with the correct linear power spectrum.  In a \LCDM\ simulation with $z_{\rm init} = 49$ and $\kNy = 4\,h~{\rm Mpc}^{-1}$, we show that the this suppression results in a 15\% power deficit near $\kNy/2$ at $z=5$, and that rescaling completely restores this power (Fig.~\ref{fig:rescaling-NL}).

We have also presented a new way to calculate second-order corrections in Lagrangian perturbation theory from direct force calculations, without the need for large Fourier transforms.  We compare our 2LPT to the actual evolution of the particle system from $z=4999$ and find excellent agreement on large scales, with differences of 1\% at $\kNy/2$ due to particle lattice anisotropy.  We also find excellent agreement with \textsc{2LPTic}, a standard FFT-based code, below $\kNy/3$.  Above that scale, both approaches have drawbacks: \textsc{2LPTic} starts to introduce transverse-mode power (as any non-PLT code would), reaching 8\% of power in transverse (i.e.~transient) modes at $\kNy$; similarly, our 2LPT suffers a scatter of 4\% due to anisotropic lattice effects at the same scale.

Finally, we have tested the impact of PLT, rescaling, and our 2LPT implementation on the matter power spectrum, halo masses, and halo two-point clustering at $z=1$ in a series of $720^3$ particle simulations initialized at $z=49$ with particle mass $4\e{10}\,h^{-1} {\rm M_\odot}$.  We compare the results to an oversampled reference simulation at 8 times the mass resolution.  While our reference configuration does not represent an absolutely converged state, increasing particle density for a fixed set of modes will necessarily converge towards the continuum limit.

The power spectrum results confirm that the combination of 2LPT and rescaling is necessary to achieve 1\% accuracy down to $\kNy$.  2LPT gives 1 to 3\% errors below $\kNy/4$, corresponding roughly to 64 particle haloes.  In other words, a simulation with 2LPT alone would need 64 times the mass resolution to achieve 1\% accuracy at $\kNy$.

We identify haloes using three halo finders and found that, with a few exceptions, our results are largely independent of the finder.  Specifically, at $z=1$ we recover the known result \citep[e.g.][]{Crocce+2006,LHuillier+2014} that 2LPT is important for the most massive haloes (5 to 10\% of haloes above 1000 particles are missing without 2LPT) and show that rescaling is necessary to correct a 5 to 10\% deficit of haloes below 500 particles.  These deficits increase at higher redshift, since the non-linearities of structure formation have not yet had time to transfer power from low $k$ to high $k$.  At all redshifts and mass ranges, 2LPT with rescaling was the best match to the reference simulation.

We analyse the halo 2PCF both in bins of absolute mass and mass rank.  The absolute mass bins show the strongest dependence on the choice of ICs (at the level of 5\%), because haloes are changing mass at constant bias, causing them to shift in or out of the mass bin.  Binning by mass rank greatly reduces this effect and lowers the dependence on the ICs to the level of 1 to 3\%.  In nearly all cases, the combination of 2LPT and rescaling produces the best match to the reference 2PCF, reaching agreement of a fraction of a percent in many cases.

Our PLT growing mode corrections are manifestly the correct way to initialize cosmological $N$-body simulations to linear order.  These corrections, in combination with rescaling and our 2LPT implementation, are crucial for recovering accurate small-scale power spectra, halo masses, and clustering.  We anticipate that our 2LPT implementation will be particularly useful for extremely large $N$-body simulations, where large FFTs are expensive.  Rescaling may be most useful for arrays of medium-resolution simulations in which substantially increasing the particle density is too expensive computationally, e.g.~when constructing covariance matrices or building cosmic emulators \citep[e.g.][]{Lawrence+2010}.

Code to generate ZA initial conditions with PLT eigenmode corrections and rescaling is available at \url{https://github.com/lgarrison/zeldovich-PLT}.

\section*{Acknowledgements}
We thank Svetlin Tassev for helpful discussions and the referee for comments that helped improve the quality of this paper.  We acknowledge use of the University of Washington N-body Shop friends-of friends code.  Some computations in this paper were run on the Odyssey cluster supported by the FAS Division of Science, Research Computing Group at Harvard University.  This work has been supported by grant AST-1313285 from the National Science Foundation and by grant DE-SC0013718 from the U.S.~Department of Energy. Some of the computations used in this study were performed on the El Gato supercomputer at the University of Arizona, supported by grant 1228509 from the National Science Foundation.

\bibliographystyle{mnras}
\bibliography{abacus_plt}

\bsp	
\label{lastpage}
\end{document}